\documentclass[12pt]{article}

\usepackage{graphicx}

\begin{document}

\begin{center}
{\large{\bf Fission Barriers of Neutron-rich and Superheavy 
Nuclei calculated with the ETFSI Method.}}$^*$\\

\vspace{0.5cm}

{\bf A. Mamdouh$^{1}$, J. M. Pearson$^{1,2}$, M. Rayet$^{1}$ and 
F. Tondeur$^{1,3}$}\\

\vspace{0.5cm}

\noindent 1. Institut d'Astronomie et d'Astrophysique, CP-226, Universit\'{e} 
Libre de Bruxelles, Boulevard du Triomphe, B-1050 Brussels, Belgium.\\
\noindent 2. D\'epartement de Physique,
Universit\'e de Montr\'eal, Montr\'eal (Qu\'ebec), H3C 3J7 Canada.\\
\noindent 3. Institut Sup\'erieur Industriel de Bruxelles.\\
\end{center}

\begin{abstract}

Using the ETFSI (extended Thomas-Fermi plus Strutinsky integral) method, 
we have calculated the fission barriers of nearly 2000 exotic nuclei, 
including all the neutron-rich nuclei 
up to $A = 318$ that are expected to be relevant to the r-process, and 
all superheavy nuclei in the vicinity of $N = 184$, with $Z \le 120$. 
Our calculations were performed with the Skyrme force SkSC4, which was 
determined in the ETFSI-1 mass fit. For proton-deficient nuclei in the
region of $N$ = 184 we find the barriers to be much higher than previously
believed, which suggests that the r-process path might continue to mass 
numbers well beyond 300. For the superheavy nuclei we typically find barrier 
heights of 6--7 MeV. \\


\noindent{$^*$Supported in part by FNRS (Belgium) and NSERC (Canada)}

\end{abstract}

\thispagestyle{empty}

\newpage

\section{Introduction}

The r-process of stellar nucleosynthesis depends crucially on the 
masses and fission barriers (among other quantities) of nuclei 
that are so neutron-rich that there is no hope of being able to 
measure them in the laboratory (see Refs. \cite{meyer94,artak99} for
reviews discussing the nuclear data required for an understanding of 
the r-process). It is thus of the greatest importance
to be able to make reliable extrapolations of these quantities away 
from the known region, relatively close to the stability 
line, out towards the neutron-drip line. Until recently the masses 
and barriers used in all studies of the r-process were calculated on 
the basis of one form or another of the liquid-drop(let) model (LDM). 
However, in an attempt to put the extrapolations on as rigorous 
a footing as possible we have developed a mass formula that is 
based entirely on microscopic 
forces, the ETFSI-1 mass formula \cite{dut,ton,pea,abo1,abo2}. Calculations of
the r-process using the ETFSI-1 masses have already been performed 
\cite{how93,ga96}, but they are incomplete in that fission had to be 
neglected, barriers not yet having been calculated in the ETFSI 
model. Of course, recourse could have been made to the extensive 
barrier calculations \cite{how80,mey89} based on the LDM that had 
been used in earlier r-process studies, but it would have been 
inconsistent to use one model for the masses (required for the 
neutron-separation energies $S_n$ and the beta-decay energies 
$Q_\beta$), and another for the barriers.

Here we remedy this deficiency by presenting the results of ETFSI-method 
calculations of the fission barriers of all of the nearly 2000 nuclei 
lying in the region of the ($N$,$Z$) plane shown in Fig. 1. This region
covers the range $84 \le Z \le 120$, and
extends from moderately neutron-deficient to extremely neutron-rich nuclei,
including thereby not only a large fraction of the nuclei whose barriers 
are known experimentally, but also all nuclei with $A \le 318$ 
whose barriers can reasonably be expected to be relevant to the r-process. 
Towards the upper limit of our range of $Z$ values the r-process path lies
at much higher values of $A$ (if it has not already been terminated by 
fission), and we restrict ourselves to nuclei lying
close to the stability line in the long-expected ``island" of stability
that is now becoming experimentally accessible \cite{oga99,nin99,ogb99}. 

The ETFSI method is a high-speed approximation to the 
Skyrme-Hartree-Fock (SHF) method, with pairing correlations 
generated by a $\delta$-function force, treated in the 
usual BCS approach (with blocking). There are two parts to the 
total energy calculated by the ETFSI method, the first consisting 
of a purely semi-classical approximation to the SHF method, the 
full fourth-order extended Thomas-Fermi (ETF) method, while the 
second part, which is based on what we call the 
Strutinsky-integral (SI) form of the Strutinsky theorem,
constitutes an attempt to improve this approximation 
perturbatively, and in particular to restore the shell corrections 
that are missing from the ETF part. The way in which we extended 
the ETFSI method, developed originally for the calculation of 
ground-state energies, to the large-scale calculation of fission 
barriers is described in Ref. \cite{fiss1}, which should be 
consulted for all details of our calculational methods. However, it is
important to recall here that our fission paths are optimized with respect
to the the elongation parameter $c$, the necking parameter $h$,
and the asymmetry parameter $\tilde{\alpha}$ (see Ref. \cite{fiss1}, and in
particular its App. A, for the definition of these quantities, and a 
description of the nuclear shapes allowed by our parametrization in the case of
fission; note that as in that reference we are assuming axial symmetry). 

In Ref. \cite{fiss1}
we calculated all the barriers that have been measured in nuclei with 
$Z \ge 81$. The comparison of our results with the available data showed that
if the primary, i,e., highest, barrier is lower than 10 MeV the error never
exceeds 1.4 MeV, with either sign being possible, while for primary barrier 
heights lying between 10 and 15 MeV we probably overestimate the height 
by about 1.5 MeV; barrier heights in excess of 15 MeV are overestimated by 
2 MeV or more. Since the primary barrier is the only one that is really
relevant to the r-process we see that our method
is sufficiently accurate for the calculation of barriers that are low enough
to be of interest to the r-process (see Section 3 below). The
calculations here are performed in exactly the same way, and in particular we
use the same force parameters, set SkSC4, which, it
should be noted, were determined entirely by the mass fit 
\cite{abo1}, and have in no way been modified for the 
barrier calculations.

 \begin{figure}
 \rotatebox{90}{\resizebox{!}{\hsize}{\includegraphics{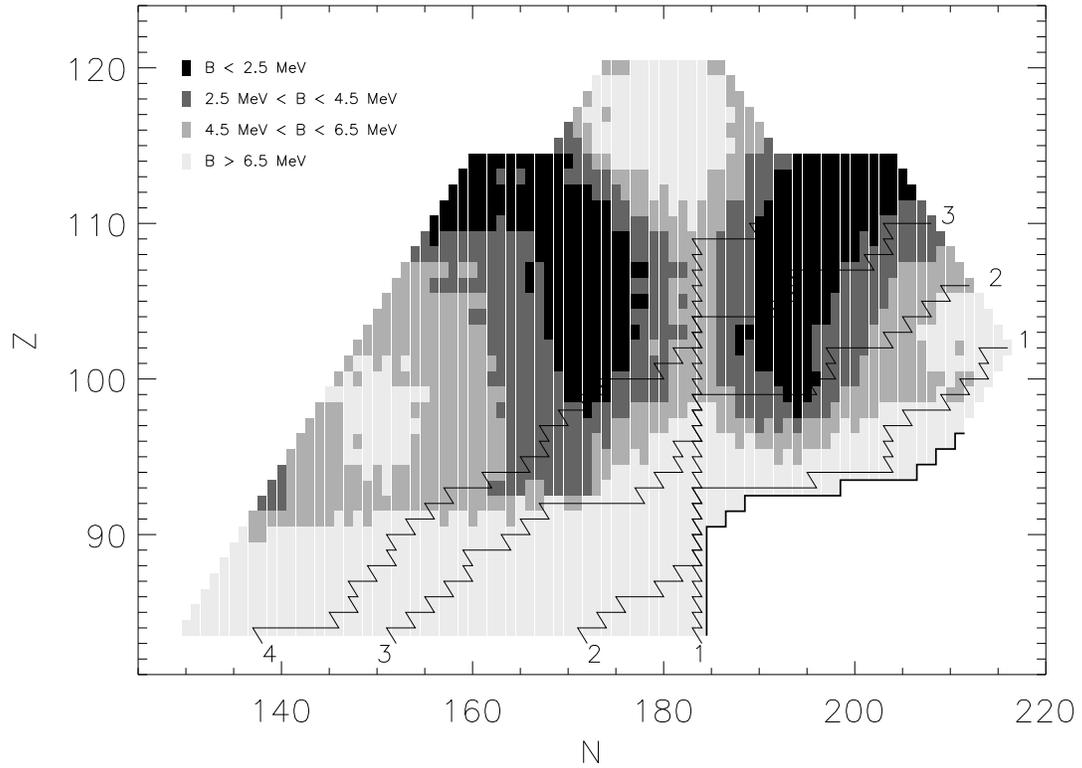}}}
 \caption{Schematic representation in the ($N$,$Z$) plane of the 
 primary fission barriers displayed in Table \ref{barriers}. The energy bins 
 are described in the figure. Thin solid lines represent r-process paths 
 for constant values (in MeV) of $S_a$ (defined in text). The
 neutron-drip line is shown by the thick solid line}
  
 \end{figure}

\section{Results}

We summarize the results of our calculations in Fig. 1, and in Table \ref{barriers}, 
where for each nucleus we present the primary barrier.
In order to keep its size within reasonable bounds, Table \ref{barriers} does not 
show the three deformation parameters of the barriers, $c$, $h$, 
and $\tilde{\alpha}$. However, these parameters 
can be found on our web site (http://www-astro.ulb.ac.be/ ),
which gives also the value of the secondary (i.e., second-highest)
barrier, when present, as well as its deformation parameters. As mentioned
in Ref. \cite{fiss1}, it is possible to distinguish without ambiguity 
between ``inner" and ``outer"  barriers from their value of $c$, 
each of the two highest barriers belonging to a 
different category. A typical situation is illustrated in Fig. 2 for the 
uranium isotope chain. For smaller values of $Z$ the separation between
primary and secondary barriers, expressed in terms of $c$, tends to 
increase, and the external barrier becomes the primary one for all
isotopes, as it is already in Fig. 2 for large neutron excesses
(this can be explained by the fact that nuclei with smaller fissility
need higher deformation to fission).
For larger $Z$ this separation generally decreases and the
inner barrier is always the highest, while the outer one progressively
disappears, so that for $Z > 100$ only one barrier (the inner one) is left. 
We label the primary barriers displayed in Table \ref{barriers} with the 
superscript $i$ or $o$ to show  whether they belong to the inner or to the
outer category, respectively.

The general trends of the results of Table \ref{barriers} are displayed in Fig. 1 as
well as by the curves labelled ``ETFSI-1" in Figs. 3--5, where we show 
three isotope chains, $Z$ = 84, 92, and 100, respectively (note that in
Fig. 3 the ETFSI-1 barriers for $A$ = 207--212, not included in 
Table \ref{barriers}, are taken from Ref. \cite{fiss1}). 

The curve labelled ``ETF" denotes the results we find for 
pure ETF calculations of the barriers without either shell or pairing
corrections. In these latter calculations we search anew the positions 
of the ground states and saddle points, as defined by their three 
deformation parameters, and find different
deformations for these points than obtained in the ETFSI calculation. 
We emphasize that if shell and pairing corrections had simply been subtracted 
from the ETFSI saddle-point and ground-state energies, the ETF
curve would not be at all as smooth as it is in Figs. 3--5.

Also seen on these graphs are the results of the LDM calculations
of Howard and M\"oller (labelled ``HM") \cite{how80}, and the results 
of Myers and Swiatecki (labelled ``MS") \cite{ms99}, based on 
zeroth-order Thomas-Fermi calculations \cite{ms96}. Both these latter sets 
of results include shell corrections calculated, in one way or 
another, by the Strutinsky method. Actually, Ref. \cite{ms99} gives just
a smooth formula representing the main trends of Thomas-Fermi barrier 
calculations, which we reproduce here as the curve ``MS.0". 
We have constructed the curve ``MS" ourselves, following the prescription of 
Ref. \cite{ms99}, i.e., we have added the shell corrections quoted in
that paper (see their Ref. [4]) to the ground-state energy only, 
assuming there are no shell corrections at the saddle points (this is their
``topographic" theorem).

\begin{figure}
\resizebox{\hsize}{!}{\includegraphics{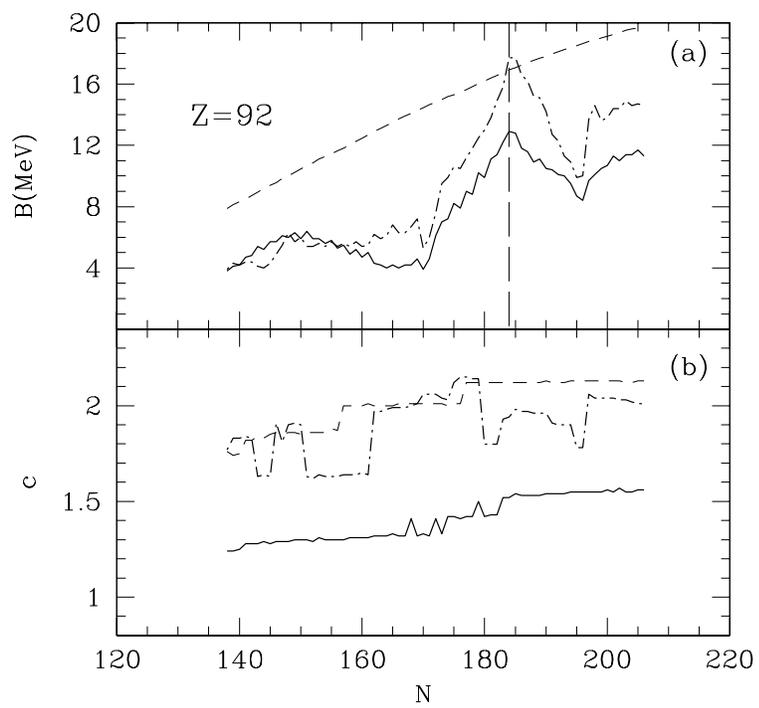}}
\caption{(a) Inner (solid curve) and outer (dot-dashed curve) 
barriers for the $Z = 92$ isotopic chain. The (single) ETF barrier is also 
shown (dashed curve). (b) The corresponding elongation parameter $c$.
}
\end{figure}

\begin{table}
\caption{Heights (in MeV) of primary barriers ($^i$ denotes inner, 
$^o$ outer -- see text)}
\label{barriers}
\begin{tabular}{cccccccccccccccccc}                                             
\hline                                                                          
$Z$&$N$&$B$&$Z$&$N$&$B$&$Z$&$N$&                                                
$B$&$Z$&$N$&$B$&$Z$&$N$&$B$&$Z$&$N$&$B$\\                                       
\hline                                                                          
 84&130&22.5$^o$& 84&160&21.5$^o$& 85&136&15.4$^o$& 85&166&21.2$^o$& 86&143&13.4$^o$& 86&173&24.6$^o$\\
   &131&21.6$^o$&   &161&21.3$^o$&   &137&15.1$^o$&   &167&21.6$^o$&   &144&13.8$^o$&   &174&24.5$^o$\\
   &132&20.9$^o$&   &162&21.5$^o$&   &138&14.9$^o$&   &168&22.5$^o$&   &145&14.2$^o$&   &175&25.3$^o$\\
   &133&20.0$^o$&   &163&21.5$^o$&   &139&14.8$^o$&   &169&22.9$^o$&   &146&14.7$^o$&   &176&25.8$^o$\\
   &134&19.4$^o$&   &164&22.1$^o$&   &140&14.7$^o$&   &170&24.4$^o$&   &147&15.2$^o$&   &177&24.8$^o$\\
   &135&18.5$^o$&   &165&22.3$^o$&   &141&14.8$^o$&   &171&25.2$^o$&   &148&16.3$^o$&   &178&26.7$^o$\\
   &136&18.3$^o$&   &166&23.5$^o$&   &142&15.2$^o$&   &172&26.4$^o$&   &149&15.1$^o$&   &179&27.2$^o$\\
   &137&17.2$^o$&   &167&23.8$^o$&   &143&15.6$^o$&   &173&26.2$^o$&   &150&16.2$^o$&   &180&28.5$^o$\\
   &138&17.6$^o$&   &168&24.8$^o$&   &144&16.3$^o$&   &174&27.0$^o$&   &151&15.7$^o$&   &181&29.7$^o$\\
   &139&16.6$^o$&   &169&25.4$^o$&   &145&16.0$^o$&   &175&27.4$^o$&   &152&16.3$^o$&   &182&31.5$^o$\\
   &140&16.8$^o$&   &170&26.9$^o$&   &146&16.7$^o$&   &176&28.1$^o$&   &153&16.4$^o$&   &183&32.9$^o$\\
   &141&17.4$^o$&   &171&27.2$^o$&   &147&16.5$^o$&   &177&28.2$^o$&   &154&16.6$^o$&   &184&33.5$^o$\\
   &142&17.8$^o$&   &172&28.9$^o$&   &148&16.7$^o$&   &178&29.1$^o$&   &155&17.3$^o$& 87&133&12.4$^o$\\
   &143&17.9$^o$&   &173&29.5$^o$&   &149&17.3$^o$&   &179&29.8$^o$&   &156&17.1$^o$&   &134&12.3$^o$\\
   &144&18.7$^o$&   &174&29.3$^o$&   &150&17.2$^o$&   &180&31.2$^o$&   &157&16.8$^o$&   &135&12.0$^o$\\
   &145&18.9$^o$&   &175&30.0$^o$&   &151&17.5$^o$&   &181&32.3$^o$&   &158&16.9$^o$&   &136&11.8$^o$\\
   &146&19.2$^o$&   &176&30.1$^o$&   &152&17.8$^o$&   &182&34.0$^o$&   &159&16.9$^o$&   &137&12.0$^o$\\
   &147&18.6$^o$&   &177&30.5$^o$&   &153&18.4$^o$&   &183&35.6$^o$&   &160&17.1$^o$&   &138&10.4$^o$\\
   &148&18.9$^o$&   &178&31.8$^o$&   &154&18.6$^o$&   &184&36.3$^o$&   &161&17.1$^o$&   &139&10.4$^o$\\
   &149&18.8$^o$&   &179&32.2$^o$&   &155&19.0$^o$& 86&132&14.8$^o$&   &162&17.5$^o$&   &140&10.4$^o$\\
   &150&19.9$^o$&   &180&34.0$^o$&   &156&19.4$^o$&   &133&13.9$^o$&   &163&17.9$^o$&   &141&10.4$^o$\\
   &151&19.3$^o$&   &181&35.5$^o$&   &157&19.1$^o$&   &134&13.4$^o$&   &164&17.7$^o$&   &142&10.4$^o$\\
   &152&20.7$^o$&   &182&37.3$^o$&   &158&19.1$^o$&   &135&12.8$^o$&   &165&18.1$^o$&   &143&10.8$^o$\\
   &153&20.9$^o$&   &183&38.7$^o$&   &159&18.9$^o$&   &136&12.7$^o$&   &166&18.7$^o$&   &144&11.1$^o$\\
   &154&21.2$^o$&   &184&39.0$^o$&   &160&19.1$^o$&   &137&12.6$^o$&   &167&19.5$^o$&   &145&11.5$^o$\\
   &155&22.0$^o$& 85&131&18.4$^o$&   &161&19.2$^o$&   &138&12.2$^o$&   &168&20.0$^o$&   &146&12.0$^o$\\
   &156&22.3$^o$&   &132&17.7$^o$&   &162&19.6$^o$&   &139&12.2$^o$&   &169&20.4$^o$&   &147&12.4$^o$\\
   &157&22.0$^o$&   &133&16.9$^o$&   &163&19.5$^o$&   &140&12.1$^o$&   &170&21.5$^o$&   &148&13.0$^o$\\
   &158&21.6$^o$&   &134&16.3$^o$&   &164&19.7$^o$&   &141&12.2$^o$&   &171&23.4$^o$&   &149&13.5$^o$\\
   &159&21.4$^o$&   &135&15.5$^o$&   &165&20.2$^o$&   &142&12.8$^o$&   &172&23.8$^o$&   &150&14.5$^o$\\
\hline                                                                          
\end{tabular}
\end{table}
\newpage                                                                        
\begin{tabular}{cccccccccccccccccc}                                             
\hline                                                                          
$Z$&$N$&$B$&$Z$&$N$&$B$&$Z$&$N$&                                                
$B$&$Z$&$N$&$B$&$Z$&$N$&$B$&$Z$&$N$&$B$\\                                       
\hline                                                                          
 87&151&14.5$^o$& 87&181&26.9$^o$& 88&160&12.8$^o$& 89&140& 8.0$^o$& 89&170&14.0$^o$& 90&151& 9.3$^o$\\
   &152&14.0$^o$&   &182&28.6$^o$&   &161&12.2$^o$&   &141& 8.1$^o$&   &171&14.5$^o$&   &152& 9.2$^o$\\
   &153&13.9$^o$&   &183&30.1$^o$&   &162&14.1$^o$&   &142& 8.3$^o$&   &172&15.4$^o$&   &153& 9.0$^o$\\
   &154&14.1$^o$&   &184&30.9$^o$&   &163&13.6$^o$&   &143& 8.6$^o$&   &173&16.4$^o$&   &154& 8.9$^o$\\
   &155&14.4$^o$& 88&134& 9.3$^o$&   &164&14.4$^o$&   &144& 8.5$^o$&   &174&17.3$^o$&   &155& 8.9$^o$\\
   &156&14.5$^o$&   &135& 9.3$^o$&   &165&14.4$^o$&   &145& 8.8$^o$&   &175&18.0$^o$&   &156& 8.3$^o$\\
   &157&14.4$^o$&   &136& 9.1$^o$&   &166&14.1$^o$&   &146& 8.7$^o$&   &176&19.4$^o$&   &157& 8.3$^o$\\
   &158&14.6$^o$&   &137& 9.0$^o$&   &167&15.1$^o$&   &147& 9.2$^o$&   &177&19.0$^o$&   &158& 8.8$^o$\\
   &159&14.4$^o$&   &138& 8.8$^o$&   &168&15.9$^o$&   &148& 9.3$^o$&   &178&18.7$^o$&   &159& 8.5$^o$\\
   &160&14.7$^o$&   &139& 8.9$^o$&   &169&15.9$^o$&   &149& 9.7$^o$&   &179&18.4$^o$&   &160& 9.7$^o$\\
   &161&15.5$^o$&   &140& 9.0$^o$&   &170&16.1$^o$&   &150& 9.7$^o$&   &180&20.1$^o$&   &161& 8.2$^o$\\
   &162&15.1$^o$&   &141& 9.4$^o$&   &171&16.6$^o$&   &151& 9.7$^o$&   &181&21.5$^o$&   &162& 8.3$^o$\\
   &163&16.0$^o$&   &142& 9.2$^o$&   &172&19.3$^o$&   &152& 9.9$^o$&   &182&22.6$^o$&   &163&10.1$^o$\\
   &164&16.4$^o$&   &143& 9.5$^o$&   &173&20.3$^o$&   &153&10.8$^o$&   &183&24.6$^o$&   &164&10.3$^o$\\
   &165&15.8$^o$&   &144& 9.5$^o$&   &174&20.2$^o$&   &154& 9.9$^o$&   &184&25.9$^o$&   &165& 8.7$^o$\\
   &166&17.2$^o$&   &145& 9.4$^o$&   &175&20.9$^o$&   &155&10.8$^o$& 90&136& 8.2$^o$&   &166&10.5$^o$\\
   &167&17.5$^o$&   &146& 9.7$^o$&   &176&21.1$^o$&   &156& 9.8$^o$&   &137& 6.4$^o$&   &167& 9.4$^o$\\
   &168&18.3$^o$&   &147& 9.9$^o$&   &177&22.1$^o$&   &157&11.0$^o$&   &138& 6.4$^o$&   &168&10.0$^o$\\
   &169&18.5$^o$&   &148&10.0$^o$&   &178&21.2$^o$&   &158& 9.7$^o$&   &139& 6.9$^o$&   &169&11.7$^o$\\
   &170&18.9$^o$&   &149&10.9$^o$&   &179&21.2$^o$&   &159& 9.6$^o$&   &140& 6.8$^o$&   &170&11.5$^o$\\
   &171&19.3$^o$&   &150&11.2$^o$&   &180&23.7$^o$&   &160& 9.5$^o$&   &141& 7.2$^o$&   &171&12.1$^o$\\
   &172&21.6$^o$&   &151&11.4$^o$&   &181&24.4$^o$&   &161& 9.9$^o$&   &142& 7.1$^o$&   &172&13.7$^o$\\
   &173&21.8$^o$&   &152&12.7$^o$&   &182&26.0$^o$&   &162&12.0$^o$&   &143& 6.8$^o$&   &173&14.0$^o$\\
   &174&23.0$^o$&   &153&12.5$^o$&   &183&27.4$^o$&   &163&10.0$^o$&   &144& 7.5$^o$&   &174&15.0$^o$\\
   &175&22.9$^o$&   &154&12.5$^o$&   &184&28.5$^o$&   &164&10.5$^o$&   &145& 7.6$^o$&   &175&15.3$^o$\\
   &176&23.4$^o$&   &155&12.1$^o$& 89&135& 8.2$^o$&   &165&10.7$^o$&   &146& 7.9$^o$&   &176&16.1$^o$\\
   &177&22.2$^o$&   &156&12.3$^o$&   &136& 8.0$^o$&   &166&11.3$^o$&   &147& 8.2$^o$&   &177&16.8$^o$\\
   &178&23.6$^o$&   &157&12.4$^o$&   &137& 7.9$^o$&   &167&11.7$^o$&   &148& 8.0$^o$&   &178&16.1$^o$\\
   &179&23.5$^o$&   &158&12.4$^o$&   &138& 7.7$^o$&   &168&12.4$^o$&   &149& 8.6$^o$&   &179&18.1$^o$\\
   &180&26.2$^o$&   &159&12.6$^o$&   &139& 8.1$^o$&   &169&13.0$^o$&   &150& 9.6$^o$&   &180&17.4$^o$\\
\hline                                                                          
\end{tabular}                                                                   
\newpage                                                                        
\begin{tabular}{cccccccccccccccccc}                                             
\hline                                                                          
$Z$&$N$&$B$&$Z$&$N$&$B$&$Z$&$N$&                                                
$B$&$Z$&$N$&$B$&$Z$&$N$&$B$&$Z$&$N$&$B$\\                                       
\hline                                                                          
 90&181&18.3$^o$& 91&163& 7.4$^o$& 92&144& 5.2$^i$& 92&174& 9.9$^o$& 93&154& 5.8$^i$& 93&184&13.8$^o$\\
   &182&20.5$^o$&   &164& 6.3$^o$&   &145& 5.7$^i$&   &175&10.6$^o$&   &155& 5.4$^i$&   &185&14.8$^o$\\
   &183&22.0$^o$&   &165& 8.5$^o$&   &146& 5.7$^i$&   &176&10.5$^o$&   &156& 5.2$^i$&   &186&14.0$^o$\\
   &184&23.5$^o$&   &166& 8.2$^o$&   &147& 6.1$^i$&   &177&11.2$^o$&   &157& 5.3$^i$&   &187&12.7$^o$\\
 91&137& 4.6$^o$&   &167& 8.4$^o$&   &148& 6.3$^o$&   &178&11.8$^o$&   &158& 4.8$^i$&   &188&12.8$^o$\\
   &138& 5.3$^o$&   &168& 8.8$^o$&   &149& 6.3$^i$&   &179&12.5$^o$&   &159& 5.2$^i$&   &189&10.9$^o$\\
   &139& 6.0$^o$&   &169& 9.2$^o$&   &150& 6.0$^o$&   &180&13.0$^o$&   &160& 4.5$^i$&   &190&11.8$^o$\\
   &140& 5.8$^o$&   &170& 7.9$^o$&   &151& 6.4$^i$&   &181&13.8$^o$&   &161& 4.8$^i$&   &191& 9.8$^o$\\
   &141& 5.9$^o$&   &171& 8.8$^o$&   &152& 5.9$^i$&   &182&14.9$^o$&   &162& 4.3$^i$&   &192& 8.9$^o$\\
   &142& 6.0$^o$&   &172&10.3$^o$&   &153& 5.9$^i$&   &183&15.8$^o$&   &163& 4.4$^i$&   &193& 8.9$^i$\\
   &143& 5.4$^i$&   &173&11.0$^o$&   &154& 5.6$^i$&   &184&17.7$^o$&   &164& 4.4$^o$&   &194& 9.1$^o$\\
   &144& 5.7$^o$&   &174&12.1$^o$&   &155& 5.8$^i$&   &185&17.7$^o$&   &165& 4.4$^o$&   &195& 9.4$^o$\\
   &145& 5.7$^o$&   &175&12.3$^o$&   &156& 5.4$^o$&   &186&16.6$^o$&   &166& 4.6$^o$&   &196& 9.4$^o$\\
   &146& 6.7$^o$&   &176&13.0$^o$&   &157& 5.6$^o$&   &187&16.2$^o$&   &167& 4.6$^o$&   &197& 9.9$^o$\\
   &147& 5.8$^i$&   &177&14.0$^o$&   &158& 5.4$^o$&   &188&15.2$^o$&   &168& 3.8$^i$&   &198&11.0$^o$\\
   &148& 7.9$^o$&   &178&14.3$^o$&   &159& 5.7$^o$& 93&139& 4.4$^i$&   &169& 3.9$^i$& 94&140& 4.2$^i$\\
   &149& 6.1$^o$&   &179&15.5$^o$&   &160& 5.4$^o$&   &140& 4.3$^i$&   &170& 3.9$^i$&   &141& 4.7$^i$\\
   &150& 7.9$^o$&   &180&14.8$^o$&   &161& 5.5$^o$&   &141& 5.0$^i$&   &171& 4.2$^i$&   &142& 4.8$^i$\\
   &151& 7.7$^o$&   &181&15.5$^o$&   &162& 6.2$^o$&   &142& 4.9$^i$&   &172& 4.2$^i$&   &143& 5.5$^i$\\
   &152& 7.6$^o$&   &182&16.9$^o$&   &163& 5.9$^o$&   &143& 5.7$^i$&   &173& 6.0$^i$&   &144& 5.4$^i$\\
   &153& 7.5$^o$&   &183&19.1$^o$&   &164& 6.1$^o$&   &144& 5.4$^i$&   &174& 7.2$^o$&   &145& 5.8$^i$\\
   &154& 7.2$^o$&   &184&20.5$^o$&   &165& 6.8$^o$&   &145& 6.0$^i$&   &175& 7.7$^o$&   &146& 5.8$^i$\\
   &155& 7.0$^o$&   &185&19.8$^o$&   &166& 6.3$^o$&   &146& 5.9$^i$&   &176& 7.9$^o$&   &147& 6.4$^i$\\
   &156& 6.9$^o$&   &186&18.7$^o$&   &167& 6.3$^o$&   &147& 6.3$^i$&   &177& 8.9$^o$&   &148& 6.2$^i$\\
   &157& 6.4$^o$& 92&138& 3.9$^o$&   &168& 6.7$^o$&   &148& 6.1$^i$&   &178& 9.0$^o$&   &149& 6.7$^i$\\
   &158& 6.5$^o$&   &139& 4.3$^o$&   &169& 7.2$^o$&   &149& 6.4$^i$&   &179&10.3$^o$&   &150& 6.4$^i$\\
   &159& 6.5$^o$&   &140& 4.2$^o$&   &170& 5.3$^o$&   &150& 6.1$^i$&   &180&10.9$^o$&   &151& 6.7$^i$\\
   &160& 6.2$^o$&   &141& 4.7$^i$&   &171& 6.0$^o$&   &151& 6.5$^i$&   &181&12.0$^o$&   &152& 6.2$^i$\\
   &161& 7.5$^o$&   &142& 4.8$^i$&   &172& 7.6$^o$&   &152& 6.1$^i$&   &182&12.4$^o$&   &153& 6.3$^i$\\
   &162& 7.5$^o$&   &143& 5.4$^i$&   &173& 9.5$^o$&   &153& 6.2$^i$&   &183&12.7$^o$&   &154& 5.9$^i$\\
\hline                                                                          
\end{tabular}                                                                   
\newpage                                                                        
\begin{tabular}{cccccccccccccccccc}                                             
\hline                                                                          
$Z$&$N$&$B$&$Z$&$N$&$B$&$Z$&$N$&                                                
$B$&$Z$&$N$&$B$&$Z$&$N$&$B$&$Z$&$N$&$B$\\                                       
\hline                                                                          
 94&155& 6.2$^i$& 94&185&11.1$^o$& 95&149& 6.9$^i$& 95&179& 7.2$^i$& 96&142& 4.9$^i$& 96&172& 3.1$^i$\\
   &156& 5.5$^i$&   &186& 9.6$^o$&   &150& 6.6$^i$&   &180& 7.4$^i$&   &143& 5.5$^i$&   &173& 4.7$^i$\\
   &157& 5.6$^i$&   &187& 9.4$^o$&   &151& 6.9$^i$&   &181& 8.4$^i$&   &144& 5.5$^i$&   &174& 4.0$^i$\\
   &158& 5.2$^i$&   &188& 8.9$^o$&   &152& 6.5$^i$&   &182& 8.5$^i$&   &145& 6.2$^i$&   &175& 5.7$^i$\\
   &159& 5.5$^i$&   &189& 8.7$^o$&   &153& 6.5$^i$&   &183& 9.5$^i$&   &146& 6.1$^i$&   &176& 5.5$^i$\\
   &160& 5.0$^i$&   &190& 8.0$^o$&   &154& 6.2$^i$&   &184& 9.4$^o$&   &147& 6.6$^i$&   &177& 6.1$^i$\\
   &161& 4.8$^i$&   &191& 8.5$^i$&   &155& 6.1$^i$&   &185& 9.0$^o$&   &148& 6.4$^i$&   &178& 5.8$^i$\\
   &162& 4.5$^i$&   &192& 7.7$^i$&   &156& 5.8$^i$&   &186& 7.8$^i$&   &149& 6.7$^i$&   &179& 6.6$^i$\\
   &163& 4.6$^i$&   &193& 7.7$^i$&   &157& 5.8$^i$&   &187& 7.9$^i$&   &150& 6.5$^i$&   &180& 6.6$^i$\\
   &164& 4.2$^i$&   &194& 7.1$^o$&   &158& 5.4$^i$&   &188& 7.0$^i$&   &151& 6.7$^i$&   &181& 7.6$^i$\\
   &165& 4.3$^i$&   &195& 8.0$^i$&   &159& 5.3$^i$&   &189& 7.6$^i$&   &152& 6.5$^i$&   &182& 7.8$^i$\\
   &166& 3.8$^i$&   &196& 8.0$^o$&   &160& 4.7$^i$&   &190& 6.9$^i$&   &153& 6.7$^i$&   &183& 8.6$^i$\\
   &167& 3.6$^i$&   &197& 8.6$^o$&   &161& 4.8$^i$&   &191& 7.3$^i$&   &154& 6.1$^i$&   &184& 8.3$^o$\\
   &168& 3.8$^i$&   &198& 9.1$^o$&   &162& 4.6$^i$&   &192& 6.4$^i$&   &155& 6.4$^i$&   &185& 8.3$^i$\\
   &169& 4.2$^i$&   &199& 9.3$^o$&   &163& 4.5$^i$&   &193& 6.5$^i$&   &156& 5.9$^i$&   &186& 6.6$^o$\\
   &170& 3.8$^i$&   &200& 9.8$^o$&   &164& 4.1$^i$&   &194& 5.8$^i$&   &157& 5.5$^i$&   &187& 7.4$^i$\\
   &171& 3.7$^i$&   &201&12.1$^o$&   &165& 4.3$^i$&   &195& 6.7$^i$&   &158& 5.0$^i$&   &188& 6.4$^i$\\
   &172& 4.1$^i$&   &202&10.5$^o$&   &166& 3.9$^i$&   &196& 6.9$^i$&   &159& 5.2$^i$&   &189& 6.7$^i$\\
   &173& 6.0$^i$&   &203&10.9$^o$&   &167& 4.0$^i$&   &197& 7.5$^i$&   &160& 4.7$^i$&   &190& 6.0$^i$\\
   &174& 5.9$^i$&   &204&11.1$^o$&   &168& 3.6$^i$&   &198& 7.6$^i$&   &161& 4.9$^i$&   &191& 6.4$^i$\\
   &175& 6.8$^o$&   &205&11.6$^o$&   &169& 3.8$^i$&   &199& 7.9$^i$&   &162& 4.5$^i$&   &192& 5.6$^i$\\
   &176& 6.9$^o$&   &206&11.5$^o$&   &170& 3.5$^i$&   &200& 8.3$^i$&   &163& 4.5$^i$&   &193& 5.6$^i$\\
   &177& 7.9$^o$& 95&141& 4.9$^i$&   &171& 3.9$^i$&   &201& 9.1$^i$&   &164& 4.2$^i$&   &194& 4.9$^i$\\
   &178& 7.9$^o$&   &142& 5.1$^i$&   &172& 3.4$^i$&   &202& 8.9$^i$&   &165& 4.2$^i$&   &195& 5.3$^i$\\
   &179& 9.0$^o$&   &143& 5.7$^i$&   &173& 5.0$^i$&   &203& 9.4$^o$&   &166& 3.7$^i$&   &196& 5.7$^i$\\
   &180& 9.6$^o$&   &144& 5.7$^i$&   &174& 5.3$^i$&   &204& 9.5$^o$&   &167& 3.6$^i$&   &197& 6.3$^i$\\
   &181&10.1$^o$&   &145& 6.3$^i$&   &175& 6.1$^i$&   &205& 9.9$^i$&   &168& 3.2$^i$&   &198& 6.7$^i$\\
   &182&10.2$^o$&   &146& 6.3$^i$&   &176& 6.0$^i$&   &206& 9.9$^o$&   &169& 3.6$^i$&   &199& 7.3$^i$\\
   &183&11.2$^o$&   &147& 6.8$^i$&   &177& 6.7$^i$&   &207&10.2$^o$&   &170& 3.1$^i$&   &200& 7.4$^i$\\
   &184&11.4$^o$&   &148& 6.5$^i$&   &178& 6.3$^o$&   &208&10.1$^o$&   &171& 3.3$^i$&   &201& 8.0$^i$\\
\hline                                                                          
\end{tabular}                                                                   
\newpage                                                                        
\begin{tabular}{cccccccccccccccccc}                                             
\hline                                                                          
$Z$&$N$&$B$&$Z$&$N$&$B$&$Z$&$N$&                                                
$B$&$Z$&$N$&$B$&$Z$&$N$&$B$&$Z$&$N$&$B$\\                                       
\hline                                                                          
 96&202& 8.0$^i$& 97&164& 4.2$^i$& 97&194& 3.8$^i$& 98&156& 5.7$^i$& 98&186& 5.0$^i$& 99&148& 6.7$^i$\\
   &203& 8.6$^i$&   &165& 4.2$^i$&   &195& 4.8$^i$&   &157& 5.8$^i$&   &187& 5.0$^i$&   &149& 7.3$^i$\\
   &204& 8.3$^i$&   &166& 3.8$^i$&   &196& 4.8$^i$&   &158& 5.2$^i$&   &188& 4.2$^i$&   &150& 7.1$^i$\\
   &205& 8.5$^i$&   &167& 3.8$^i$&   &197& 5.1$^i$&   &159& 5.2$^i$&   &189& 4.4$^i$&   &151& 7.6$^i$\\
   &206& 8.4$^i$&   &168& 3.3$^i$&   &198& 6.0$^i$&   &160& 4.8$^i$&   &190& 3.6$^i$&   &152& 7.3$^i$\\
   &207& 9.0$^i$&   &169& 3.2$^i$&   &199& 7.8$^o$&   &161& 4.9$^i$&   &191& 3.9$^i$&   &153& 6.9$^i$\\
   &208& 8.5$^i$&   &170& 3.0$^i$&   &200& 6.0$^o$&   &162& 4.5$^i$&   &192& 2.9$^i$&   &154& 6.5$^i$\\
   &209& 8.9$^i$&   &171& 2.9$^i$&   &201& 7.9$^o$&   &163& 4.3$^i$&   &193& 3.0$^i$&   &155& 6.5$^i$\\
   &210& 8.6$^i$&   &172& 2.8$^i$&   &202& 7.5$^o$&   &164& 4.0$^i$&   &194& 2.0$^i$&   &156& 6.0$^i$\\
 97&143& 6.0$^i$&   &173& 3.2$^i$&   &203& 7.9$^o$&   &165& 4.1$^i$&   &195& 3.3$^i$&   &157& 6.1$^i$\\
   &144& 5.9$^i$&   &174& 3.1$^i$&   &204& 7.7$^o$&   &166& 3.6$^i$&   &196& 3.8$^i$&   &158& 5.6$^i$\\
   &145& 6.3$^i$&   &175& 5.0$^i$&   &205& 7.8$^o$&   &167& 3.6$^i$&   &197& 4.3$^i$&   &159& 5.4$^i$\\
   &146& 6.4$^i$&   &176& 4.6$^i$&   &206& 8.0$^o$&   &168& 3.1$^i$&   &198& 5.0$^i$&   &160& 5.0$^i$\\
   &147& 6.8$^i$&   &177& 5.4$^i$&   &207& 8.4$^o$&   &169& 2.9$^i$&   &199& 5.5$^i$&   &161& 4.8$^i$\\
   &148& 6.7$^i$&   &178& 5.3$^i$&   &208& 8.0$^o$&   &170& 2.5$^i$&   &200& 5.7$^i$&   &162& 4.6$^i$\\
   &149& 7.2$^i$&   &179& 5.6$^i$&   &209& 8.1$^o$&   &171& 2.7$^i$&   &201& 6.1$^i$&   &163& 4.5$^i$\\
   &150& 6.9$^i$&   &180& 5.6$^i$&   &210& 8.1$^o$&   &172& 2.4$^i$&   &202& 6.2$^i$&   &164& 4.1$^i$\\
   &151& 7.1$^i$&   &181& 6.6$^i$&   &211& 8.4$^o$&   &173& 2.7$^i$&   &203& 6.8$^i$&   &165& 4.0$^i$\\
   &152& 6.9$^i$&   &182& 6.6$^i$& 98&144& 5.5$^i$&   &174& 2.8$^i$&   &204& 6.6$^i$&   &166& 3.6$^i$\\
   &153& 7.1$^i$&   &183& 7.3$^i$&   &145& 6.0$^i$&   &175& 4.2$^i$&   &205& 7.2$^o$&   &167& 3.6$^i$\\
   &154& 6.6$^i$&   &184& 7.5$^i$&   &146& 6.1$^i$&   &176& 4.2$^i$&   &206& 7.3$^o$&   &168& 3.0$^i$\\
   &155& 6.4$^i$&   &185& 7.1$^i$&   &147& 6.7$^i$&   &177& 5.0$^i$&   &207& 7.4$^o$&   &169& 2.9$^i$\\
   &156& 5.9$^i$&   &186& 5.8$^i$&   &148& 6.5$^i$&   &178& 5.0$^i$&   &208& 6.8$^o$&   &170& 2.3$^i$\\
   &157& 5.8$^i$&   &187& 5.3$^i$&   &149& 6.9$^i$&   &179& 5.3$^i$&   &209& 7.4$^o$&   &171& 2.3$^i$\\
   &158& 5.2$^i$&   &188& 4.5$^i$&   &150& 6.7$^i$&   &180& 5.3$^i$&   &210& 7.0$^o$&   &172& 2.0$^i$\\
   &159& 5.3$^i$&   &189& 4.7$^i$&   &151& 6.9$^i$&   &181& 6.1$^i$&   &211& 7.3$^o$&   &173& 2.4$^i$\\
   &160& 4.9$^i$&   &190& 3.8$^i$&   &152& 6.7$^i$&   &182& 6.1$^i$&   &212& 7.1$^o$&   &174& 2.1$^i$\\
   &161& 5.0$^i$&   &191& 5.5$^i$&   &153& 7.0$^i$&   &183& 7.0$^i$& 99&145& 6.6$^i$&   &175& 3.6$^i$\\
   &162& 4.5$^i$&   &192& 4.4$^i$&   &154& 6.2$^i$&   &184& 6.8$^i$&   &146& 6.5$^i$&   &176& 3.9$^i$\\
   &163& 4.5$^i$&   &193& 4.7$^i$&   &155& 6.3$^i$&   &185& 6.5$^i$&   &147& 7.1$^i$&   &177& 4.4$^i$\\
\hline                                                                          
\end{tabular}                                                                   
\newpage                                                                        
\begin{tabular}{cccccccccccccccccc}                                             
\hline                                                                          
$Z$&$N$&$B$&$Z$&$N$&$B$&$Z$&$N$&                                                
$B$&$Z$&$N$&$B$&$Z$&$N$&$B$&$Z$&$N$&$B$\\                                       
\hline                                                                          
 99&178& 3.6$^i$& 99&208& 6.3$^i$&100&170& 2.2$^i$&100&200& 4.0$^i$&101&162& 4.5$^i$&101&192& 1.6$^i$\\
   &179& 4.4$^i$&   &209& 6.6$^i$&   &171& 2.2$^i$&   &201& 4.7$^i$&   &163& 4.3$^i$&   &193& 1.2$^i$\\
   &180& 4.4$^i$&   &210& 6.2$^i$&   &172& 1.8$^i$&   &202& 4.7$^i$&   &164& 4.0$^i$&   &194& 1.7$^i$\\
   &181& 5.1$^i$&   &211& 6.4$^i$&   &173& 2.3$^i$&   &203& 5.1$^i$&   &165& 3.6$^i$&   &195& 1.9$^i$\\
   &182& 5.0$^i$&   &212& 6.5$^i$&   &174& 1.8$^i$&   &204& 5.3$^i$&   &166& 3.2$^i$&   &196& 2.8$^i$\\
   &183& 6.2$^i$&   &213& 6.8$^i$&   &175& 3.1$^i$&   &205& 5.4$^i$&   &167& 3.1$^i$&   &197& 3.1$^i$\\
   &184& 5.9$^i$&100&146& 5.9$^i$&   &176& 2.6$^i$&   &206& 5.9$^i$&   &168& 2.6$^i$&   &198& 3.6$^i$\\
   &185& 5.3$^i$&   &147& 6.5$^i$&   &177& 4.0$^i$&   &207& 5.8$^i$&   &169& 2.2$^i$&   &199& 4.2$^i$\\
   &186& 4.8$^i$&   &148& 6.3$^i$&   &178& 3.3$^i$&   &208& 6.0$^i$&   &170& 1.9$^i$&   &200& 4.3$^i$\\
   &187& 4.7$^i$&   &149& 6.8$^i$&   &179& 4.2$^i$&   &209& 6.4$^i$&   &171& 1.7$^i$&   &201& 4.7$^i$\\
   &188& 3.7$^i$&   &150& 6.7$^i$&   &180& 4.3$^i$&   &210& 7.3$^i$&   &172& 1.4$^i$&   &202& 5.0$^i$\\
   &189& 4.0$^i$&   &151& 6.8$^i$&   &181& 5.2$^i$&   &211& 6.3$^i$&   &173& 1.8$^i$&   &203& 5.3$^i$\\
   &190& 3.2$^i$&   &152& 6.3$^i$&   &182& 5.1$^i$&   &212& 6.4$^i$&   &174& 1.3$^i$&   &204& 5.7$^i$\\
   &191& 3.6$^i$&   &153& 6.4$^i$&   &183& 6.3$^i$&   &213& 8.3$^i$&   &175& 2.4$^i$&   &205& 5.6$^i$\\
   &192& 2.6$^i$&   &154& 5.8$^i$&   &184& 6.0$^i$&   &214& 8.4$^i$&   &176& 1.9$^i$&   &206& 6.4$^i$\\
   &193& 2.4$^i$&   &155& 6.0$^i$&   &185& 5.5$^i$&101&147& 6.4$^i$&   &177& 3.2$^i$&   &207& 6.1$^i$\\
   &194& 2.0$^i$&   &156& 5.5$^i$&   &186& 4.3$^i$&   &148& 6.5$^i$&   &178& 3.0$^i$&   &208& 7.0$^i$\\
   &195& 2.1$^i$&   &157& 5.6$^i$&   &187& 4.0$^i$&   &149& 6.6$^i$&   &179& 3.9$^i$&   &209& 7.5$^i$\\
   &196& 2.5$^i$&   &158& 5.1$^i$&   &188& 3.4$^i$&   &150& 6.4$^i$&   &180& 3.4$^i$&   &210& 7.3$^i$\\
   &197& 2.8$^i$&   &159& 5.2$^i$&   &189& 3.5$^i$&   &151& 6.8$^i$&   &181& 4.7$^i$&   &211& 6.6$^i$\\
   &198& 3.6$^i$&   &160& 4.8$^i$&   &190& 2.7$^i$&   &152& 6.4$^i$&   &182& 4.5$^i$&   &212& 6.5$^i$\\
   &199& 4.5$^i$&   &161& 4.7$^i$&   &191& 2.9$^i$&   &153& 6.5$^i$&   &183& 5.7$^i$&   &213& 8.5$^i$\\
   &200& 4.6$^i$&   &162& 4.4$^i$&   &192& 2.1$^i$&   &154& 6.0$^i$&   &184& 5.3$^i$&   &214& 7.8$^i$\\
   &201& 5.1$^i$&   &163& 4.1$^i$&   &193& 1.9$^i$&   &155& 6.2$^i$&   &185& 5.3$^i$&   &215& 8.2$^i$\\
   &202& 5.1$^i$&   &164& 3.8$^i$&   &194& 1.6$^i$&   &156& 5.7$^i$&   &186& 4.2$^i$&102&148& 5.8$^i$\\
   &203& 5.4$^i$&   &165& 3.8$^i$&   &195& 1.7$^i$&   &157& 6.1$^i$&   &187& 3.4$^i$&   &149& 6.0$^i$\\
   &204& 5.4$^i$&   &166& 3.3$^i$&   &196& 2.1$^i$&   &158& 5.3$^i$&   &188& 2.7$^i$&   &150& 5.8$^i$\\
   &205& 5.7$^i$&   &167& 3.2$^i$&   &197& 3.1$^i$&   &159& 5.3$^i$&   &189& 3.0$^i$&   &151& 6.2$^i$\\
   &206& 5.8$^i$&   &168& 2.7$^i$&   &198& 3.4$^i$&   &160& 5.0$^i$&   &190& 2.2$^i$&   &152& 5.7$^i$\\
   &207& 6.3$^i$&   &169& 2.6$^i$&   &199& 4.1$^i$&   &161& 4.8$^i$&   &191& 2.4$^i$&   &153& 5.8$^i$\\
\hline                                                                          
\end{tabular}                                                                   
\newpage                                                                        
\begin{tabular}{cccccccccccccccccc}                                             
\hline                                                                          
$Z$&$N$&$B$&$Z$&$N$&$B$&$Z$&$N$&                                                
$B$&$Z$&$N$&$B$&$Z$&$N$&$B$&$Z$&$N$&$B$\\                                       
\hline                                                                          
102&154& 5.6$^i$&102&184& 5.6$^i$&102&214& 7.6$^i$&103&176& 1.2$^i$&103&206& 6.1$^i$&104&170& 1.3$^i$\\
   &155& 5.8$^i$&   &185& 5.5$^i$&   &215& 8.0$^i$&   &177& 2.4$^i$&   &207& 6.6$^i$&   &171& 1.5$^i$\\
   &156& 5.3$^i$&   &186& 4.1$^i$&   &216& 7.7$^i$&   &178& 2.5$^i$&   &208& 6.7$^i$&   &172&  .8$^i$\\
   &157& 5.5$^i$&   &187& 4.2$^i$&103&149& 6.0$^i$&   &179& 3.7$^i$&   &209& 7.0$^i$&   &173& 1.3$^i$\\
   &158& 5.0$^i$&   &188& 2.4$^i$&   &150& 5.9$^i$&   &180& 3.3$^i$&   &210& 7.1$^i$&   &174& 1.0$^i$\\
   &159& 5.1$^i$&   &189& 2.5$^i$&   &151& 6.3$^i$&   &181& 4.2$^i$&   &211& 7.6$^i$&   &175& 1.5$^i$\\
   &160& 4.8$^i$&   &190& 1.7$^i$&   &152& 5.9$^i$&   &182& 4.1$^i$&   &212& 7.5$^i$&   &176& 1.6$^i$\\
   &161& 4.6$^i$&   &191& 2.2$^i$&   &153& 6.1$^i$&   &183& 5.3$^i$&   &213& 7.9$^i$&   &177& 2.7$^i$\\
   &162& 4.3$^i$&   &192& 1.2$^i$&   &154& 5.8$^i$&   &184& 4.9$^i$&   &214& 7.7$^i$&   &178& 2.7$^i$\\
   &163& 4.0$^i$&   &193&  .9$^i$&   &155& 6.1$^i$&   &185& 4.8$^i$&   &215& 8.1$^i$&   &179& 3.9$^i$\\
   &164& 3.7$^i$&   &194& 1.1$^i$&   &156& 5.6$^i$&   &186& 4.2$^i$&104&150& 5.3$^i$&   &180& 3.7$^i$\\
   &165& 3.6$^i$&   &195& 2.2$^i$&   &157& 5.7$^i$&   &187& 4.0$^i$&   &151& 5.6$^i$&   &181& 5.0$^i$\\
   &166& 3.0$^i$&   &196& 2.6$^i$&   &158& 5.2$^i$&   &188& 1.7$^i$&   &152& 5.3$^i$&   &182& 4.7$^i$\\
   &167& 2.5$^i$&   &197& 2.8$^i$&   &159& 5.4$^i$&   &189& 2.1$^i$&   &153& 5.6$^i$&   &183& 5.8$^i$\\
   &168& 2.5$^i$&   &198& 3.3$^i$&   &160& 4.9$^i$&   &190& 1.2$^i$&   &154& 5.2$^i$&   &184& 5.6$^i$\\
   &169& 2.3$^i$&   &199& 3.8$^i$&   &161& 4.9$^i$&   &191& 1.3$^i$&   &155& 5.3$^i$&   &185& 4.8$^i$\\
   &170& 1.7$^i$&   &200& 3.9$^i$&   &162& 4.4$^i$&   &192&  .3$^i$&   &156& 5.0$^i$&   &186& 4.0$^i$\\
   &171& 1.7$^i$&   &201& 4.4$^i$&   &163& 4.0$^i$&   &193&  .3$^i$&   &157& 5.0$^i$&   &187& 3.9$^i$\\
   &172& 1.3$^i$&   &202& 4.6$^i$&   &164& 3.7$^i$&   &194& 1.9$^i$&   &158& 4.7$^i$&   &188& 2.9$^i$\\
   &173& 1.7$^i$&   &203& 5.0$^i$&   &165& 3.7$^i$&   &195& 2.2$^i$&   &159& 4.6$^i$&   &189& 2.9$^i$\\
   &174& 1.3$^i$&   &204& 5.4$^i$&   &166& 3.1$^i$&   &196& 2.6$^i$&   &160& 4.4$^i$&   &190& 1.9$^i$\\
   &175& 1.7$^i$&   &205& 5.7$^i$&   &167& 2.9$^i$&   &197& 2.9$^i$&   &161& 4.4$^i$&   &191& 1.9$^i$\\
   &176& 1.7$^i$&   &206& 6.0$^i$&   &168& 2.4$^i$&   &198& 3.3$^i$&   &162& 4.0$^i$&   &192&  .4$^i$\\
   &177& 3.0$^i$&   &207& 5.7$^i$&   &169& 2.2$^i$&   &199& 3.8$^i$&   &163& 3.6$^i$&   &193& 1.0$^i$\\
   &178& 2.8$^i$&   &208& 6.5$^i$&   &170& 1.6$^i$&   &200& 4.0$^i$&   &164& 3.3$^i$&   &194& 1.5$^i$\\
   &179& 4.6$^i$&   &209& 7.0$^i$&   &171& 1.4$^i$&   &201& 4.4$^i$&   &165& 3.0$^i$&   &195& 1.9$^i$\\
   &180& 3.8$^i$&   &210& 7.1$^i$&   &172& 1.1$^i$&   &202& 4.7$^i$&   &166& 2.5$^i$&   &196& 2.1$^i$\\
   &181& 4.8$^i$&   &211& 6.2$^i$&   &173& 1.2$^i$&   &203& 4.9$^i$&   &167& 2.6$^i$&   &197& 2.3$^i$\\
   &182& 4.9$^i$&   &212& 7.3$^i$&   &174&  .9$^i$&   &204& 5.3$^i$&   &168& 1.9$^i$&   &198& 2.9$^i$\\
   &183& 6.0$^i$&   &213& 7.8$^i$&   &175& 1.3$^i$&   &205& 5.8$^i$&   &169& 2.0$^i$&   &199& 3.3$^i$\\
\hline                                                                          
\end{tabular}                                                                   
\newpage                                                                        
\begin{tabular}{cccccccccccccccccc}                                             
\hline                                                                          
$Z$&$N$&$B$&$Z$&$N$&$B$&$Z$&$N$&                                                
$B$&$Z$&$N$&$B$&$Z$&$N$&$B$&$Z$&$N$&$B$\\                                       
\hline                                                                          
104&200& 3.4$^i$&105&166& 2.8$^i$&105&196& 1.9$^i$&106&164& 3.3$^i$&106&194& 1.0$^i$&107&164& 3.7$^i$\\
   &201& 4.1$^i$&   &167& 2.8$^i$&   &197& 2.1$^i$&   &165& 3.0$^i$&   &195& 1.3$^i$&   &165& 3.7$^i$\\
   &202& 4.3$^i$&   &168& 1.8$^i$&   &198& 2.7$^i$&   &166& 2.3$^i$&   &196& 1.5$^i$&   &166& 2.4$^i$\\
   &203& 4.5$^i$&   &169& 2.0$^i$&   &199& 2.9$^i$&   &167& 2.5$^i$&   &197& 2.1$^i$&   &167& 2.7$^i$\\
   &204& 5.0$^i$&   &170& 1.4$^i$&   &200& 3.3$^i$&   &168& 1.4$^i$&   &198& 2.3$^i$&   &168& 1.9$^i$\\
   &205& 5.4$^i$&   &171& 1.2$^i$&   &201& 3.9$^i$&   &169& 1.7$^i$&   &199& 2.6$^i$&   &169& 2.1$^i$\\
   &206& 5.6$^i$&   &172& 1.9$^i$&   &202& 4.0$^i$&   &170& 1.0$^i$&   &200& 2.9$^i$&   &170& 1.3$^i$\\
   &207& 6.1$^i$&   &173& 1.9$^i$&   &203& 4.4$^i$&   &171& 1.2$^i$&   &201& 3.4$^i$&   &171& 1.4$^i$\\
   &208& 6.2$^i$&   &174& 1.5$^i$&   &204& 5.1$^i$&   &172&  .7$^i$&   &202& 3.7$^i$&   &172& 1.1$^i$\\
   &209& 6.6$^i$&   &175& 1.9$^i$&   &205& 5.6$^i$&   &173& 1.4$^i$&   &203& 4.2$^i$&   &173& 1.5$^i$\\
   &210& 6.6$^i$&   &176& 1.5$^i$&   &206& 5.6$^i$&   &174& 1.2$^i$&   &204& 4.6$^i$&   &174& 1.4$^i$\\
   &211& 7.0$^i$&   &177& 2.0$^i$&   &207& 6.0$^i$&   &175& 1.8$^i$&   &205& 4.8$^i$&   &175& 1.8$^i$\\
   &212& 7.0$^i$&   &178& 2.1$^i$&   &208& 6.0$^i$&   &176& 1.6$^i$&   &206& 5.0$^i$&   &176& 1.6$^i$\\
   &213& 7.5$^i$&   &179& 3.0$^i$&   &209& 6.5$^i$&   &177& 2.7$^i$&   &207& 5.5$^i$&   &177& 2.3$^i$\\
   &214& 7.3$^i$&   &180& 3.3$^i$&   &210& 6.5$^i$&   &178& 2.5$^i$&   &208& 5.5$^i$&   &178& 2.2$^i$\\
105&151& 5.7$^i$&   &181& 4.5$^i$&   &211& 6.9$^i$&   &179& 3.5$^i$&   &209& 6.0$^i$&   &179& 3.0$^i$\\
   &152& 5.5$^i$&   &182& 4.2$^i$&   &212& 6.9$^i$&   &180& 3.7$^i$&   &210& 6.0$^i$&   &180& 3.1$^i$\\
   &153& 5.6$^i$&   &183& 5.3$^i$&   &213& 7.2$^i$&   &181& 5.1$^i$&   &211& 6.3$^i$&   &181& 4.3$^i$\\
   &154& 5.2$^i$&   &184& 4.6$^i$&106&152& 4.7$^i$&   &182& 4.8$^i$&   &212& 6.2$^i$&   &182& 4.0$^i$\\
   &155& 5.3$^i$&   &185& 4.7$^i$&   &153& 4.9$^i$&   &183& 5.7$^i$&107&153& 4.8$^i$&   &183& 5.2$^i$\\
   &156& 5.1$^i$&   &186& 3.7$^i$&   &154& 4.6$^i$&   &184& 5.1$^i$&   &154& 4.6$^i$&   &184& 4.6$^i$\\
   &157& 5.2$^i$&   &187& 3.8$^i$&   &155& 4.7$^i$&   &185& 4.8$^i$&   &155& 4.8$^i$&   &185& 4.6$^i$\\
   &158& 4.9$^i$&   &188& 2.7$^i$&   &156& 4.3$^i$&   &186& 3.8$^i$&   &156& 4.5$^i$&   &186& 3.5$^i$\\
   &159& 4.9$^i$&   &189& 2.9$^i$&   &157& 4.4$^i$&   &187& 4.2$^i$&   &157& 4.5$^i$&   &187& 4.0$^i$\\
   &160& 4.6$^i$&   &190& 1.8$^i$&   &158& 4.2$^i$&   &188& 3.4$^i$&   &158& 4.4$^i$&   &188& 2.7$^i$\\
   &161& 4.5$^i$&   &191& 1.7$^i$&   &159& 4.3$^i$&   &189& 2.9$^i$&   &159& 4.6$^i$&   &189& 2.8$^i$\\
   &162& 4.2$^i$&   &192&  .7$^i$&   &160& 3.9$^i$&   &190& 2.0$^i$&   &160& 4.5$^i$&   &190& 2.0$^i$\\
   &163& 4.1$^i$&   &193& 1.0$^i$&   &161& 3.8$^i$&   &191& 1.9$^i$&   &161& 4.3$^i$&   &191& 1.6$^i$\\
   &164& 3.7$^i$&   &194& 1.5$^i$&   &162& 3.7$^i$&   &192&  .8$^i$&   &162& 3.9$^i$&   &192&  .7$^i$\\
   &165& 3.1$^i$&   &195& 1.8$^i$&   &163& 3.8$^i$&   &193&  .5$^i$&   &163& 4.0$^i$&   &193&  .2$^i$\\
\hline                                                                          
\end{tabular}                                                                   
\newpage                                                                        
\begin{tabular}{cccccccccccccccccc}                                             
\hline                                                                          
$Z$&$N$&$B$&$Z$&$N$&$B$&$Z$&$N$&                                                
$B$&$Z$&$N$&$B$&$Z$&$N$&$B$&$Z$&$N$&$B$\\                                       
\hline                                                                          
107&194&  .9$^i$&108&166& 2.5$^i$&108&196& 1.2$^i$&109&170& 1.6$^i$&109&200& 2.0$^i$&110&176& 3.1$^i$\\
   &195& 1.3$^i$&   &167& 2.1$^i$&   &197& 1.2$^i$&   &171& 1.7$^i$&   &201& 2.5$^i$&   &177& 4.2$^i$\\
   &196& 1.6$^i$&   &168& 1.9$^i$&   &198& 1.6$^i$&   &172& 1.4$^i$&   &202& 2.2$^i$&   &178& 4.1$^i$\\
   &197& 1.8$^i$&   &169& 1.6$^i$&   &199& 1.9$^i$&   &173& 1.8$^i$&   &203& 3.0$^i$&   &179& 5.1$^i$\\
   &198& 2.2$^i$&   &170& 1.3$^i$&   &200& 2.3$^i$&   &174& 1.6$^i$&   &204& 3.1$^i$&   &180& 5.0$^i$\\
   &199& 2.6$^i$&   &171& 1.4$^i$&   &201& 2.7$^i$&   &175& 2.1$^i$&   &205& 3.6$^i$&   &181& 6.0$^i$\\
   &200& 2.8$^i$&   &172& 1.2$^i$&   &202& 2.9$^i$&   &176& 1.7$^i$&   &206& 3.6$^i$&   &182& 5.6$^i$\\
   &201& 3.3$^i$&   &173& 1.6$^i$&   &203& 3.4$^i$&   &177& 2.7$^i$&   &207& 4.0$^i$&   &183& 6.6$^i$\\
   &202& 3.5$^i$&   &174& 1.3$^i$&   &204& 3.5$^i$&   &178& 2.6$^i$&   &208& 4.1$^i$&   &184& 6.0$^i$\\
   &203& 4.1$^i$&   &175& 2.1$^i$&   &205& 3.9$^i$&   &179& 4.4$^i$&   &209& 4.3$^i$&   &185& 5.9$^i$\\
   &204& 4.3$^i$&   &176& 1.9$^i$&   &206& 4.1$^i$&   &180& 4.2$^i$&110&156& 1.9$^i$&   &186& 5.3$^i$\\
   &205& 4.8$^i$&   &177& 2.7$^i$&   &207& 4.3$^i$&   &181& 5.0$^i$&   &157& 2.2$^i$&   &187& 4.4$^i$\\
   &206& 4.8$^i$&   &178& 2.5$^i$&   &208& 4.4$^i$&   &182& 4.9$^i$&   &158& 2.1$^i$&   &188& 3.7$^i$\\
   &207& 5.3$^i$&   &179& 4.0$^i$&   &209& 4.8$^i$&   &183& 5.8$^i$&   &159& 2.3$^i$&   &189& 3.2$^i$\\
   &208& 5.3$^i$&   &180& 3.8$^i$&   &210& 5.0$^i$&   &184& 5.5$^i$&   &160& 1.8$^i$&   &190& 2.3$^i$\\
   &209& 5.6$^i$&   &181& 5.0$^i$&109&155& 3.1$^i$&   &185& 5.3$^i$&   &161& 2.2$^i$&   &191& 2.1$^i$\\
   &210& 5.8$^i$&   &182& 4.9$^i$&   &156& 2.4$^i$&   &186& 4.4$^i$&   &162& 2.2$^i$&   &192& 1.1$^i$\\
   &211& 6.2$^i$&   &183& 6.2$^i$&   &157& 3.4$^i$&   &187& 4.4$^i$&   &163& 2.8$^i$&   &193&  .8$^i$\\
108&154& 3.6$^i$&   &184& 5.6$^i$&   &158& 2.8$^i$&   &188& 3.3$^i$&   &164& 1.8$^i$&   &194&  .2$^i$\\
   &155& 3.9$^i$&   &185& 5.3$^i$&   &159& 3.1$^i$&   &189& 3.4$^i$&   &165& 2.3$^i$&   &195&  .1$^i$\\
   &156& 3.7$^i$&   &186& 4.2$^i$&   &160& 3.2$^i$&   &190& 2.2$^i$&   &166& 2.1$^i$&   &196& -.6$^i$\\
   &157& 3.9$^i$&   &187& 4.4$^i$&   &161& 3.4$^i$&   &191& 2.0$^i$&   &167& 2.0$^i$&   &197&  .5$^i$\\
   &158& 3.5$^i$&   &188& 3.1$^i$&   &162& 3.3$^i$&   &192& 1.2$^i$&   &168& 1.7$^i$&   &198&  .9$^i$\\
   &159& 3.5$^i$&   &189& 3.2$^i$&   &163& 3.0$^i$&   &193&  .5$^i$&   &169& 2.0$^i$&   &199& 1.2$^i$\\
   &160& 3.4$^i$&   &190& 2.2$^i$&   &164& 2.7$^i$&   &194& -.6$^i$&   &170& 1.6$^i$&   &200& 1.4$^i$\\
   &161& 3.5$^i$&   &191& 1.9$^i$&   &165& 3.1$^i$&   &195& -.5$^i$&   &171& 1.8$^i$&   &201& 1.8$^i$\\
   &162& 3.4$^i$&   &192&  .9$^i$&   &166& 2.6$^i$&   &196&  .7$^i$&   &172& 1.5$^i$&   &202& 1.8$^i$\\
   &163& 3.6$^i$&   &193&  .3$^i$&   &167& 2.3$^i$&   &197&  .8$^i$&   &173& 2.3$^i$&   &203& 2.5$^i$\\
   &164& 2.6$^i$&   &194&  .1$^i$&   &168& 2.1$^i$&   &198& 1.4$^i$&   &174& 1.7$^i$&   &204& 2.5$^i$\\
   &165& 2.9$^i$&   &195&  .9$^i$&   &169& 1.8$^i$&   &199& 1.8$^i$&   &175& 2.6$^i$&   &205& 3.1$^i$\\
\hline                                                                          
\end{tabular}                                                                   
\newpage                                                                        
\begin{tabular}{cccccccccccccccccc}                                             
\hline                                                                          
$Z$&$N$&$B$&$Z$&$N$&$B$&$Z$&$N$&                                                
$B$&$Z$&$N$&$B$&$Z$&$N$&$B$&$Z$&$N$&$B$\\                                       
\hline                                                                          
110&206& 3.0$^i$&111&184& 6.4$^i$&112&164& 1.6$^i$&112&194&  .7$^i$&113&176& 5.7$^i$&114&160& 1.7$^i$\\
   &207& 3.3$^i$&   &185& 5.7$^i$&   &165& 2.0$^i$&   &195& -.5$^i$&   &177& 6.3$^i$&   &161& 2.1$^i$\\
   &208& 3.5$^i$&   &186& 5.0$^i$&   &166& 1.9$^i$&   &196& -.2$^i$&   &178& 6.3$^i$&   &162& 2.2$^i$\\
111&157&  .9$^i$&   &187& 4.4$^i$&   &167& 1.8$^i$&   &197& -.3$^i$&   &179& 7.2$^i$&   &163& 2.0$^i$\\
   &158& 1.2$^i$&   &188& 3.8$^i$&   &168& 1.6$^i$&   &198& -.1$^i$&   &180& 7.3$^i$&   &164& 1.8$^i$\\
   &159& 1.7$^i$&   &189& 3.4$^i$&   &169& 1.9$^i$&   &199&  .1$^i$&   &181& 7.4$^i$&   &165& 2.3$^i$\\
   &160& 2.2$^i$&   &190& 2.6$^i$&   &170& 1.8$^i$&   &200&  .4$^i$&   &182& 7.1$^i$&   &166& 1.9$^i$\\
   &161& 1.5$^i$&   &191& 2.2$^i$&   &171& 2.2$^i$&   &201&  .8$^i$&   &183& 7.4$^i$&   &167& 2.2$^i$\\
   &162& 1.7$^i$&   &192& 1.5$^i$&   &172& 2.2$^i$&   &202&  .9$^i$&   &184& 6.7$^i$&   &168& 2.2$^i$\\
   &163& 2.3$^i$&   &193&  .9$^i$&   &173& 2.7$^i$&   &203& 1.4$^i$&   &185& 6.3$^i$&   &169& 2.4$^i$\\
   &164& 2.0$^i$&   &194&  .4$^i$&   &174& 3.6$^i$&   &204& 1.5$^i$&   &186& 5.9$^i$&   &170& 2.4$^i$\\
   &165& 2.4$^i$&   &195&  .0$^i$&   &175& 4.3$^i$&   &205& 2.0$^i$&   &187& 5.5$^i$&   &171& 2.7$^i$\\
   &166& 1.6$^i$&   &196& -.6$^i$&   &176& 4.8$^i$&   &206& 1.9$^i$&   &188& 4.7$^i$&   &172& 4.1$^i$\\
   &167& 2.1$^i$&   &197&  .4$^i$&   &177& 5.6$^i$&113&159& 1.6$^i$&   &189& 4.2$^i$&   &173& 4.8$^i$\\
   &168& 2.0$^i$&   &198&  .4$^i$&   &178& 5.8$^i$&   &160& 1.9$^i$&   &190& 3.4$^i$&   &174& 6.1$^i$\\
   &169& 1.9$^i$&   &199&  .7$^i$&   &179& 6.5$^i$&   &161& 1.9$^i$&   &191& 2.7$^i$&   &175& 6.7$^i$\\
   &170& 1.7$^i$&   &200& 1.0$^i$&   &180& 6.4$^i$&   &162& 2.1$^i$&   &192& 2.1$^i$&   &176& 6.6$^i$\\
   &171& 2.1$^i$&   &201& 1.3$^i$&   &181& 7.1$^i$&   &163& 2.6$^i$&   &193& 1.6$^i$&   &177& 7.3$^i$\\
   &172& 1.8$^i$&   &202& 1.3$^i$&   &182& 6.9$^i$&   &164& 1.8$^i$&   &194&  .9$^i$&   &178& 7.2$^i$\\
   &173& 2.4$^i$&   &203& 2.0$^i$&   &183& 6.9$^i$&   &165& 2.5$^i$&   &195&  .6$^i$&   &179& 7.6$^i$\\
   &174& 2.4$^i$&   &204& 2.0$^i$&   &184& 6.5$^i$&   &166& 2.4$^i$&   &196&  .4$^i$&   &180& 7.5$^i$\\
   &175& 4.1$^i$&   &205& 2.5$^i$&   &185& 6.3$^i$&   &167& 2.3$^i$&   &197&  .1$^i$&   &181& 8.2$^i$\\
   &176& 3.8$^i$&   &206& 2.5$^i$&   &186& 5.7$^i$&   &168& 2.2$^i$&   &198&  .1$^i$&   &182& 7.8$^i$\\
   &177& 4.9$^i$&   &207& 2.9$^i$&   &187& 5.2$^i$&   &169& 2.4$^i$&   &199& -.8$^i$&   &183& 8.1$^i$\\
   &178& 4.8$^i$&112&158&  .8$^i$&   &188& 4.5$^i$&   &170& 2.5$^i$&   &200& -.6$^i$&   &184& 7.5$^i$\\
   &179& 5.6$^i$&   &159& 1.1$^i$&   &189& 3.8$^i$&   &171& 3.2$^i$&   &201&  .5$^i$&   &185& 7.2$^i$\\
   &180& 5.5$^i$&   &160& 1.3$^i$&   &190& 3.1$^i$&   &172& 2.8$^i$&   &202&-1.3$^i$&   &186& 6.5$^i$\\
   &181& 6.5$^i$&   &161& 1.6$^i$&   &191& 2.5$^i$&   &173& 3.2$^i$&   &203& 1.1$^i$&   &187& 6.1$^i$\\
   &182& 6.2$^i$&   &162& 2.0$^i$&   &192& 1.9$^i$&   &174& 4.4$^i$&   &204& 1.0$^i$&   &188& 5.3$^i$\\
   &183& 6.9$^i$&   &163& 2.3$^i$&   &193& 1.2$^i$&   &175& 5.0$^i$&   &205& 1.4$^i$&   &189& 4.4$^i$\\
\hline                                                                          
\end{tabular}                                                                   
\newpage                                                                        
\begin{tabular}{cccccccccccccccccc}                                             
\hline                                                                          
$Z$&$N$&$B$&$Z$&$N$&$B$&$Z$&$N$&                                                
$B$&$Z$&$N$&$B$&$Z$&$N$&$B$&$Z$&$N$&$B$\\                                       
\hline                                                                          
114&190& 3.9$^i$&115&175& 7.7$^i$&116&174& 6.9$^i$&117&175& 6.6$^i$&118&178& 7.0$^i$&119&183& 7.8$^i$\\
   &191& 3.5$^i$&   &176& 7.8$^i$&   &175& 7.5$^i$&   &176& 6.8$^i$&   &179& 7.6$^i$&   &184& 7.4$^i$\\
   &192& 3.0$^i$&   &177& 7.3$^i$&   &176& 6.5$^i$&   &177& 7.4$^i$&   &180& 7.4$^i$&   &185& 7.3$^i$\\
   &193& 2.5$^i$&   &178& 7.4$^i$&   &177& 7.0$^i$&   &178& 7.3$^i$&   &181& 7.9$^i$&   &186& 6.6$^i$\\
   &194& 1.8$^i$&   &179& 8.1$^i$&   &178& 7.2$^i$&   &179& 8.0$^i$&   &182& 7.7$^i$&   &187& 6.0$^i$\\
   &195& 1.1$^i$&   &180& 7.9$^i$&   &179& 7.7$^i$&   &180& 7.8$^i$&   &183& 7.7$^i$&120&174& 5.7$^i$\\
   &196&  .8$^i$&   &181& 8.5$^i$&   &180& 7.6$^i$&   &181& 8.4$^i$&   &184& 7.4$^i$&   &175& 6.1$^i$\\
   &197&  .4$^i$&   &182& 8.4$^i$&   &181& 8.2$^i$&   &182& 8.1$^i$&   &185& 7.1$^i$&   &176& 6.2$^i$\\
   &198&  .4$^i$&   &183& 8.7$^i$&   &182& 7.8$^i$&   &183& 8.5$^i$&   &186& 6.5$^i$&   &177& 6.7$^i$\\
   &199&  .2$^i$&   &184& 8.2$^i$&   &183& 8.3$^i$&   &184& 7.9$^i$&   &187& 5.9$^i$&   &178& 6.6$^i$\\
   &200&  .1$^i$&   &185& 7.7$^i$&   &184& 7.7$^i$&   &185& 7.9$^i$&   &188& 5.6$^i$&   &179& 7.1$^i$\\
   &201&  .1$^i$&   &186& 7.3$^i$&   &185& 7.4$^i$&   &186& 7.1$^i$&119&173& 7.1$^i$&   &180& 6.8$^i$\\
   &202&  .1$^i$&   &187& 7.0$^i$&   &186& 6.9$^i$&   &187& 6.5$^i$&   &174& 7.2$^i$&   &181& 7.2$^i$\\
   &203&  .2$^i$&   &188& 6.0$^i$&   &187& 6.1$^i$&   &188& 6.3$^i$&   &175& 6.8$^i$&   &182& 7.2$^i$\\
   &204&  .0$^i$&   &189& 5.7$^i$&   &188& 5.8$^i$&   &189& 5.7$^i$&   &176& 6.7$^i$&   &183& 7.2$^i$\\
115&169& 2.6$^i$&   &190& 4.9$^i$&   &189& 5.1$^i$&118&172& 6.0$^i$&   &177& 7.2$^i$&   &184& 6.8$^i$\\
   &170& 2.6$^i$&   &191& 4.4$^i$&   &190& 4.5$^i$&   &173& 7.1$^i$&   &178& 7.0$^i$&   &185& 6.2$^i$\\
   &171& 3.6$^i$&116&170& 3.9$^i$&117&171& 5.4$^i$&   &174& 6.8$^i$&   &179& 7.5$^i$&   &186& 5.8$^i$\\
   &172& 6.0$^i$&   &171& 4.9$^i$&   &172& 6.6$^i$&   &175& 7.4$^i$&   &180& 7.4$^i$&   &   &    \\
   &173& 6.9$^i$&   &172& 6.0$^i$&   &173& 6.5$^i$&   &176& 6.6$^i$&   &181& 8.0$^i$&   &   &    \\
   &174& 7.2$^i$&   &173& 6.8$^i$&   &174& 6.3$^i$&   &177& 7.1$^i$&   &182& 7.7$^i$&   &   &    \\
\hline                                                                          
\end{tabular}                                                                   
\newpage                                                                        
                                                                  
\section{Discussion}

\subsection{ Comparison with experiment}

The solid circles in Figs. 3--5 denote the few experimental 
barrier heights that have been measured for these isotope chains \cite{smi93}. 
In general, all three of the calculations ``ETFSI-1", ``HM", and ``MS"
are in good agreement with the data, although in the case of 
$Z = 84$ the ``ETFSI-1" barriers are between 2 and 4 MeV 
too high. This discrepancy reflects the general tendency noted above and in 
Ref. \cite{fiss1}: we overestimate all barriers higher than 15 MeV
by 2 MeV or more. Possible reasons for this are discussed in 
Ref. \cite{fiss1}; another point not mentioned there is that barriers
will be much more sensitive than masses to the droplet-model curvature 
coefficient $a_{cv}$ \cite{ms69} corresponding to the Skyrme force. Thus it is 
conceivable that
despite the good mass fit our $a_{cv}$ is sufficiently overestimated to
lead to serious errors at extreme deformations. (The Thomas-Fermi calculations
of Ref. \cite{ms96} encountered the same problem, but resolved it by adjusting
the so-called ``congruence energy"; we stress that we do not make use of this
feature in our calculations, and indeed if we had done so then the good 
agreement with experiment that we find for the lower barriers would have been 
destroyed.)
In any case, these large errors are of no practical consequence for the 
r-process, since nuclei with such high barriers will be effectively 
stable against fission in this context. On the other hand, for all barriers 
lower than 15 MeV the  ETFSI-1 results never disagree with experiment by more 
than 1.7 MeV, and usually by much less: for $Z \ge 86$ the rms error is 
717 KeV, while for $Z \ge 88$ it is as small as 698 KeV (this includes 
a few measured nuclei that are given in Table \ref{coefficients} of 
Ref. \cite{fiss1} but not here).

\begin{figure}
\resizebox{\hsize}{!}{\includegraphics{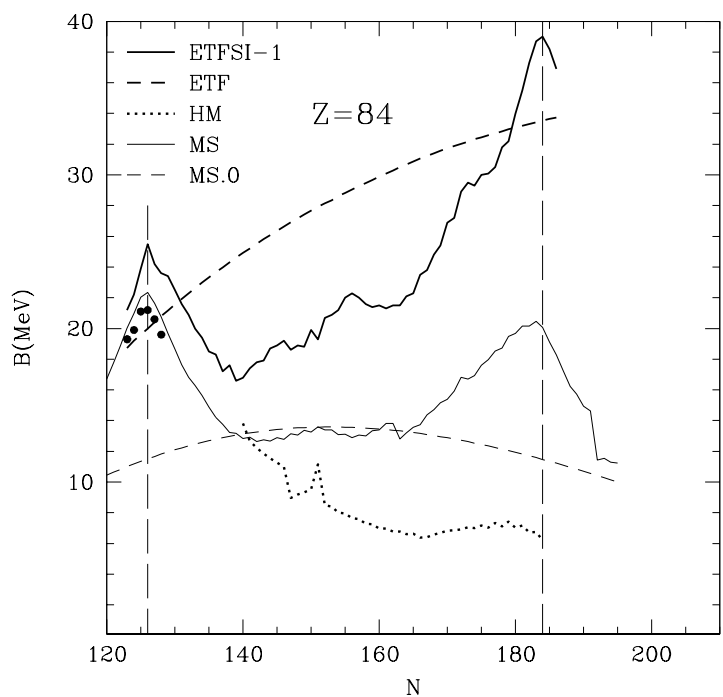}}
\caption{Primary fission barriers for the $Z=84$ isotopic chain, 
calculated with the various models considered in text. Neutron closed
shells are indicated by vertical dashed lines, experimental values by solid
circles.
}
\end{figure}

\subsection{Highly neutron-rich nuclei}

While all three methods agree reasonably well over the narrow
experimentally known region, they give widely different
extrapolations to the highly neutron-rich region. In particular, the 
ETFSI-1 calculations predict exceptionally high barriers in the region of 
$N$ = 184 for the lower values of $Z$, although as $Z$ increases, i.e., 
as the stability line is approached, these
barriers become lower, and the various sets of calculations tend to
converge. Actually, a similar trend in the vicinity
of $N=184$ can be observed in the case of the MS calculations, although
on a much lesser scale, while the HM calculations predict no enhancement
of barrier heights at all close to $N = 184$. Because of our
tendency to overestimate barriers that are high anyway, the trend in the
ETFSI-1 results shown in Figs. 3--5 may be somewhat exaggerated. However, 
we can estimate these errors by referring to Table \ref{barriers} of 
Ref. \cite{fiss1}, and we find that our calculated 
barrier for nucleus $Z=84$, $N=184$ is probably no
more than 12 MeV too high, while that of $Z=92$, $N=184$ is only 2 MeV
too high at the most. Thus the tendency we have reported for ETFSI-1 is 
at least qualitatively real, and it is certainly stronger than in
the MS case.

\begin{figure}
\resizebox{\hsize}{!}{\includegraphics{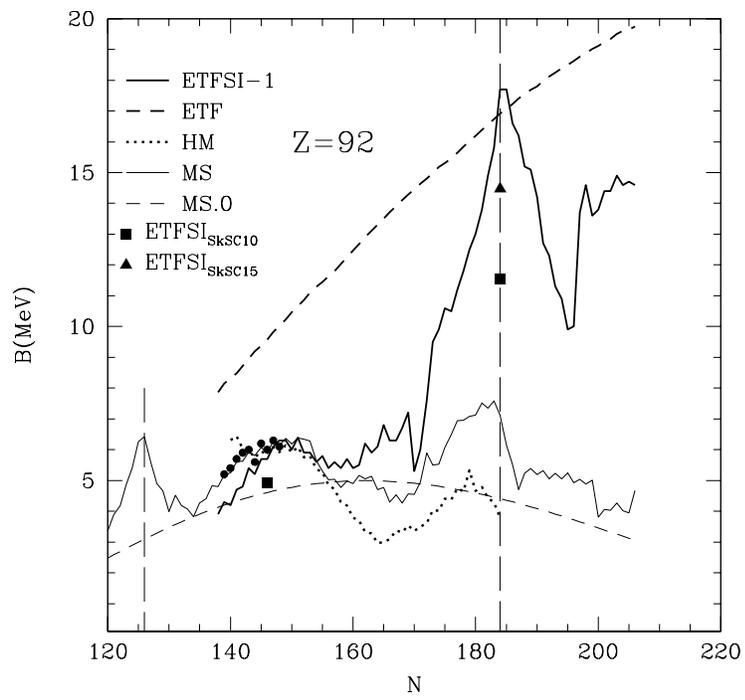}}
\caption{Same as Fig. 3 for $Z=92$. Also shown on this figure are
some results for forces SkSC10 and SkSC15.
}
\end{figure}

The rapid fall-off beyond $N = 184$ for all three isotope chains is 
indicative of a shell effect. In fact, the behaviour of the
ETFSI-1 barriers in the vicinity of $N = 184$ must be at least partially 
related to the fact that the force SkSC4 leads to a strong magic gap at 
$N$ = 184 for proton-deficient nuclei, the gap becoming much smaller as 
the stability line is approached with increasing $Z$. This point is 
illustrated in Fig. 6, where we show the variation of the gap 
$\Delta = S_{2n}(Z, N=186) - S_{2n}(Z, N=184)$ with $Z$. (It must be
recalled that we have calculated pairing in the BCS approximation, and
that with the more realistic Bogolyubov treatment much weaker neutron 
shell gaps will be found in general for large neutron excesses \cite{dob95}.
However, unpublished results of Dobaczewski, discussed by Pearson {\it
et al.} \cite{pea95}, show that this ``Bogolyubov quenching" of shell
gaps is much less pronounced for the $N = 184$ magic number, and we
therefore neglect it here.) We also show
in Fig. 6 the variation of this same gap for the FRDM (``finite-range
droplet model") mass formula \cite{frdm}. Since the shell corrections
of the MS calculations consist entirely of those of the FRDM (applied
exclusively to the ground state) we can understand why the MS barriers peak
much less strongly near $N = 184$.  

\begin{figure}
\resizebox{\hsize}{!}{\includegraphics{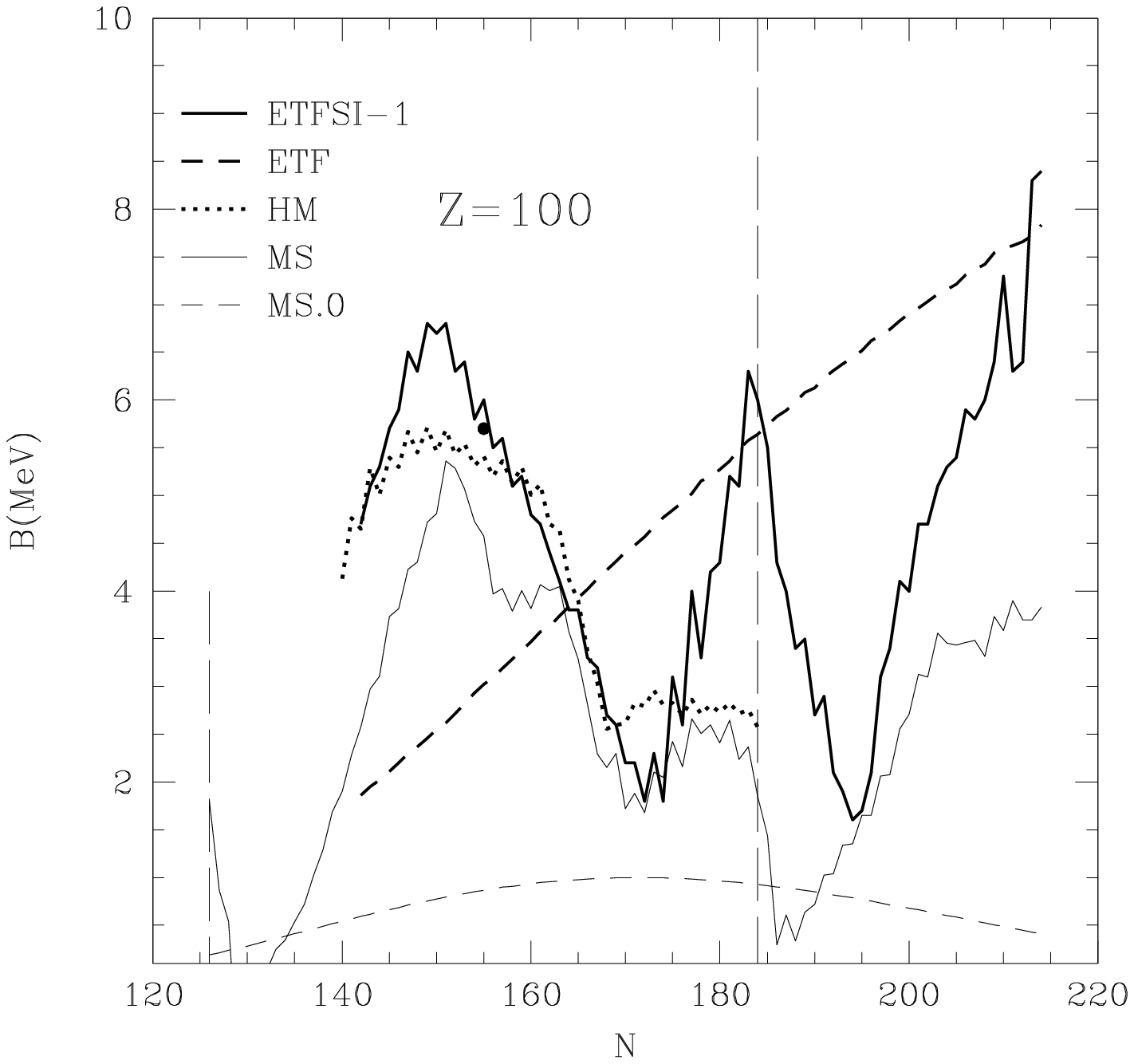}}
\caption{Same as Fig. 3 for $Z=100$
}
\end{figure}

\noindent {\it Sensitivity of barrier heights to symmetry coefficient}

However, the very high barriers found for ETFSI-1 near $N = 184$ cannot
be the result entirely of shell effects, since the pure ETF calculations
displayed in Figs. 3--5 show that the macroscopic part plays a role also, 
there being a
steady increase of barrier height with $N$, for constant $Z$. Referring to
Table \ref{coefficients}, one sees in fact that the barrier heights of these 
highly neutron-rich nuclei are strongly anti-correlated with the symmetry 
coefficient $J$ of the model in question, or equivalently, strongly correlated 
with the surface-symmetry coefficient $a_{ss}$, defined by 
\begin{equation}
\label{1}
a_{ss} = (2L/K_v)a_{sf} - 9J^2/4Q \quad ,
\end{equation}
where $L$ is the density-symmetry coefficient, $K_v$ the incompressibility,
$a_{sf}$ the surface coefficient, and $Q$ the surface-symmetry stiffness
coefficient \cite{ms69}. The anti-correlation between $J$ and $a_{ss}$
holds for all entries in Table \ref{coefficients}, and is, in fact, a 
quite general property of all models that have been fitted to nuclear masses, 
whether they are of the droplet type or are based on microscopic 
forces \cite{fcp80,fcp81}.

This result can be easily understood in terms of the following gross
approximation to the droplet model, which holds best close to the stability 
line,
\begin{equation}
\label{2}
e = a_v + a_{sf}A^{-1/3} + (J + a_{ss}A^{-1/3})I^2 + 
a_{coul}Z^2A^{-4/3} + \dots \quad ,
\end{equation}
where $e$ is the energy per nucleon and $I = (N-Z)/A$.
We see now that for ground-state masses of nuclei relatively close to 
the stability line an
increase in $J$ can be roughly compensated by a decrease in $a_{ss}$ over a 
large range of values of $A$. 
Now the term in $a_{ss}A^{-1/3}$ is really a surface term, so we can
write the fissility parameter as
\begin{equation}
\label{3}
x = \frac{a_{coul}Z^2}{2a_{sf}(I)A} \quad ,
\end{equation}
where
\begin{equation}\label{4}
a_{sf}(I) = a_{sf} + a_{ss}I^2 \quad .
\end{equation}
Thus, while a decrease in J will be compensated by an increase in $a_{ss}$
as far as ground-state masses are concerned (at least for nuclei relatively
close to the stability line), the result will be an increased barrier height
for nuclei of large neutron excess $I$.

\begin{figure}
\resizebox{\hsize}{!}{\includegraphics{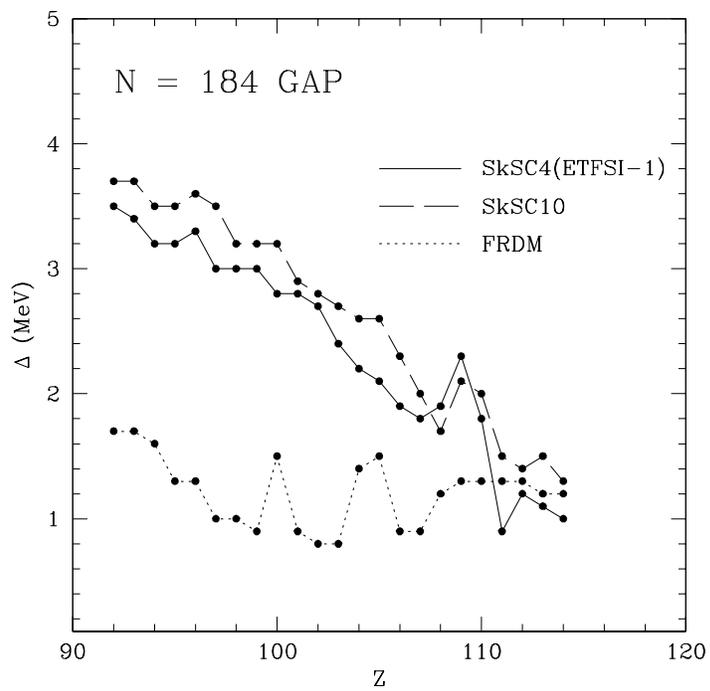}}
\caption{The $N$ = 184 pairing gap as a function of $Z$ for
SkSC4, SkSC10 and FRDM.
}
\end{figure}

This interpretation of our very high barriers as being at least in part
related to macroscopic symmetry properties is strengthened by recalculating
the barrier of U$_{184}$ with force SkSC10, a force that has been fitted in
exactly the same way as force SkSC4, except that it is constrained to $J$ = 
32 MeV \cite{opp1}, rather than 27 MeV: it will be 
seen from Fig. 4 that this barrier is lowered 
by more than 6 MeV. At the same time, for a nucleus much closer to the
stability line, U$_{146}$, the barrier is lowered by only 1 MeV on replacing
force SkSC4 by SkSC10, as we would expect for a symmetry-related effect, and
we see that the agreement with the experimental barriers that we found in
Ref. \cite{fiss1} is more or less $J$-independent. 
However, our new barrier for U$_{184}$ is still considerably higher than that
given by the MS calculation, even though the latter has almost the same 
value for $J$ (32.65 MeV). This suggests that shell effects are still playing a
major role in our high barriers, an indication that is confirmed in Fig. 6, 
where it will be seen that the shell
effects for our two Skyrme forces are more or less the same.

Nevertheless, it is clear that there is a strong macroscopic symmetry effect, 
at least within the ETFSI framework, and it is obviously essential that we
tie down much better the value of $J$ (and thus of $a_{ss}$). 
Now the rms error of the mass fit for
force SkSC10 is much worse than for the ETFSI-1 force, SkSC4 ($\sigma$
= 0.893 MeV rather than 0.736 MeV), and we now find that the {\it highest} 
value of $J$ for which acceptable mass fits can be found in the ETFSI 
framework is 28 MeV.
On the other hand, this is the {\it lowest} value
for which the known stability of neutron matter can be assured, so we now 
regard 28 MeV as our best value for $J$ \cite{pn99}. Recent HF calculations 
\cite{ton99} reach an identical conclusion, and it now seems 
that 28 MeV is a quite robust value within the very general framework of
Skyrme forces. This conclusion stands in contradiction with
the value of $J$ = 32.73 MeV given by the FRDM \cite{frdm}, the best 
droplet-model fit to masses; the almost identical value 
adopted in the zeroth-order Thomas-Fermi calculations of MS is presumably
dictated by the requirements of self-consistency, the shell corrections of
these calculations coming from the FRDM.
Ref. \cite{pn99} suggests one way in which the
FRDM could lead to a spuriously large value of $J$, but in any case we have
no option in an ETFSI-model calculation of barriers but to choose lower
values of $J$ if we wish to simultaneously fit masses.

\begin{table}[h]
\caption{Volume- and surface-symmetry coefficients of 
the forces and models used in this paper (see text)}
\label{coefficients}
\renewcommand{\tabcolsep}{1pc} 
\begin{tabular}{cccccc}
\hline
&SkSC4&MS&HM& SkSC10&SkSC15\\
\hline
$J$ (MeV) &27.0 &32.65 & 36.5 & 32.0 & 28.0\\
$L$ (MeV) &-9.29 &49.9& 100 & 55.82 & 6.73\\
$K_v$ (MeV) & 234.7 & 234 & 240 & 235.8 & 234.9\\
$a_{sf}$ (MeV) &17.7 & 18.63 & 20.76 & 18.11 & 17.78\\
$Q$ (MeV) &75$^*$ &35.4  &17.0 & 34& 56\\ 
$a_{ss}$ (MeV)  &-23.3 &-59.9 & -159&-59.2&-30.5\\
Ref. & \cite{far99} & \cite{ms96}& \cite{pm99}& \cite{far99}& \cite{far99}\\
\hline
\end{tabular}\\[2pt]
$^*$This newly calculated value replaces the one given in Ref. \cite{abo1}.
\end{table}

The question now arises as to how much our barriers would be lowered if we 
changed $J$ from the value of 27 MeV corresponding to the
force SkSC4 with which they were calculated (and which
leads to a unphysical collapse of neutron matter) to our newly preferred 
value of 28 MeV. We have accordingly constructed a new force, SkSC15, that has 
been fitted to the same mass data as SkSC4, but under the constraint of $J$ = 
28 MeV, rather than 27 MeV \cite{pn99}. Repeating the calculation of U$_{184}$, 
we see from Fig. 4 that the barrier is lowered by 3.2 MeV, which still leaves 
it much higher than the MS and HM values. Moreover, this is an extreme
case, and closer to the stability line the effect of changing $J$ by 1 MeV 
will be negligible.

\noindent{\it The r-process}

The fact that the barriers we find for very neutron-rich nuclei are much
higher than hitherto believed has significant
implications for the r-process, some of which have been presented in 
Ref.\cite{pdoye}. Here we confine ourselves to a few general remarks.

For fission to occur on a timescale 
that is short compared to the beta-decay lifetimes of the nuclei found on
the r-process path, and the r-process path thereby terminated, the nucleus in 
question will have to be excited close to or over the top of the barrier. Such 
excitation of a nucleus during the r-process can occur as the result of
either neutron capture (neutron-induced fission) or beta-decay (beta-delayed 
fission), so for rapid fission to occur the $S_n$ or the $Q_{\beta}$ 
(of the parent nucleus), respectively, must be close to or higher than the
height of the primary barrier.
Now Fig. 1 displays r-process paths corresponding
to different values of $S_a$, defined as half the two-neutron separation
energy $S_{2n}$, as well as the neutron-drip line (all calculated with the
ETFSI-1 mass formula). With typical values of $S_a$ lying between 1 and 2 MeV,
we see from this figure that neutron-induced fission will certainly not occur
in the r-process below $A = 318$, and probably not below much higher values.
On the other hand, referring to the ETFSI-1 mass tables \cite{abo2} for
the relevant $Q_{\beta}$'s, we find that beta-delayed fission will
certainly be possible for some r-process nuclei with $A < 300$, but the
extent to which this occurs depends on the beta-decay strength function 
of the precursor being sufficiently concentrated towards the upper end of 
the spectrum. But some fraction of r-process nuclei will always escape 
beta-delayed fission, and with neutron-induced fission no longer being
operative we now have to entertain the possibility of the r-process path
extending to values of $A$ considerably in excess of 300. This conclusion
is not vitiated by a possible overestimation of our barrier heights, the
limits of which are discussed in the previous sub-section.

\subsection{Superheavy nuclei}

We have already pointed out how our barriers in the vicinity of $N = 184$,
while very high on the r-process path, 
become lower as $Z$ increases, but even as the stability line is approached
they become higher again, confirming all previous expectations of an 
``island" of stability in this region. (Actually, isofar as one speaks 
of the ``valley of stability", it would be more appropriate to speak of 
a ``basin of stability" than of an ``island".) Thus
for nuclides in the range $112 \le Z \le 120$ and $177 \le N \le 186$
our calculated barriers are always at least 5 MeV, and occasionally
nearly 9 MeV, high, assuring thereby a large measure of
stability, at least with respect to fission. As for the recently discovered
superheavies, $^{289}114$ \cite{ogb99} and $^{293}118$ \cite{nin99}, 
our calculated barrier heights 
are 6.7 and 7.4 MeV, respectively. The barriers of these nuclei have not yet
been measured, but the very fact that they are stable enough to have
been observed at all constitutes a qualitative verification of the high 
barriers that our calculations predict.

As for the extent to which we agree with other microscopic 
calculations in this region
we note that in a SHF calculation on $^{288}112$ \'Cwiok {\it et al.} find a 
barrier height of about 6.5 MeV \cite{cwiok96}, as compared with our value 
of 4.8 MeV. This difference is not very large, but their barrier is a double 
one, with both peaks having about the same height, whereas we have a single, 
inner, barrier. All in all, it looks very much as though this nucleus is more
deformable in our calculation than in that of Ref. \cite{cwiok96}, despite
the fact that the force they use, SLy7 \cite{sly}, has a lower surface
coefficient $a_{sf}$ than our own SkSC4 (17.0 MeV rather than 17.7 MeV). 
The reason could be that the force SLy7 \cite{sly} has a low
effective mass $M^*$, 0.69$M$, which will lead to too low a density of
single-particle states, and thus an unrealistically high resistance to
deformation. (The fact that Ref. \cite{cwiok96} imposes
reflection symmetry while we do not cannot be a factor, since in our
calculations this nucleus turns out to be reflection-symmetric anyway.)

The only other microscopic calculations in this region of which we are 
aware are those of Berger {\it et al.} \cite{ber96}, who use the HF-Bogolyubov 
(HFB) method with the finite-range Gogny force. They find a barrier height of
10 MeV for $^{294}112$, of 11 MeV for $^{298}114$, and of 7 MeV for
$^{306}118$; these results
are to be compared with our own values of 6.9, 7.5, and 5.6 MeV, respectively.
The consistently higher barriers obtained by Ref. \cite{ber96} could be
accounted for in the same way that we have suggested for the case of
Ref. \cite{cwiok96}: too low an effective mass for the Gogny force.
However, an additional contribution could now come from the surface 
coefficient, since this is higher for the Gogny force (20.1 MeV \cite{cot77})
than for SkSC4. (The fact that Ref. \cite{ber96} imposes reflection symmetry 
while we do not can be a factor only in the case of $^{306}118$, since we 
find the other two nuclei to be reflection-symmetric anyway.)

As for the MS procedure of Ref. \cite{ms99}, it cannot be applied to 
nuclei with $Z > 112$, since this corresponds to a limiting fissility 
beyond which no prescription is given in that paper for calculating the 
macroscopic barriers. Of course, the original TF method 
\cite{ms96} on which Ref. \cite{ms99} was based is just as applicable in the 
superheavy region as elsewhere, but no such calculations seem to have been 
published.

\section{Conclusions}

We have extended the ETFSI-1 mass formula, based on the Skyrme force SkSC4, to 
the calculation of the fission
barriers of all nuclei that can be expected to play a role in the r-process
of nucleosynthesis; we recall that the force SkSC4 gives an excellent fit
to the mass data, and to the known primary-barrier heights that are lower than
15 MeV. The results that we present here 
are radically different from the only other
such calculations that have been made \cite{how80,ms99}, in that we obtain
much higher barriers for proton-deficient nuclei in the region of $N$ = 184,
a consequence of which is that the r-process path might 
continue to mass numbers considerably in excess of 300 before being brought to 
a halt by neutron-induced fission. In view of the importance of this result
we summarize here the reasons why we believe that our calculations are
essentially correct. Our high barriers on the r-path are
related both to the shell effects associated with our Skyrme force, and to our
much lower value of the symmetry coefficient $J$ (27 MeV as opposed to 36.5
MeV for HM \cite{how80} and 32.65 MeV for MS \cite{ms99}); we now examine 
each of these points in turn.

Our shell effects appear to be fairly robust within the ETFSI framework, being
difficult to change significantly while maintaining the fit to the mass data.
At the same time, we recall that in extrapolating far from the known region of
the nuclear chart out to the highly neutron-rich region that is relevant to the
r-process, the isospin dependence of the ETFSI spin-orbit field conforms well
to the predictions of relativistic mean-field theory \cite{ons97,np98}, adding
thereby to our confidence in the reliability of the extrapolation.

As for the value of the symmetry coefficient $J$, we have found, after the
completion of these calculations, that 28 MeV would have been 
a better value than 27 MeV, essentially because it allows us to avoid an 
unphysical collapse of neutron matter, while maintaining a high-quality mass
fit. A complete new mass fit, ETFSI-2, is currently 
being undertaken with the constraint $J$ = 28 MeV, and in 
principle when this is completed we should repeat the calculations of this 
paper with the new force determined by this fit. However, we have shown here
that the maximum effect of the change of force will be to lower the barriers
of the most neutron-rich nuclei by about 3 MeV, and that for nuclei closer to
the stability line the effect will be correspondingly smaller. In any case, one
can be reasonably sure that the ETFSI-2 fission barriers would lie considerably
closer to the ETFSI-1 barriers presented here than to the barriers of either
MS \cite{ms99} or HM \cite{how80}.

We have also extended our barrier calculations of nuclei with
$N$ in the vicinity of 184 down to the stability line, exploring thereby the
much studied ``island" of stability. Here our barriers are 
in reasonable agreement with other microscopic 
calculations that have been made in this region: this constitutes an additional
check on the overall validity of our calculations, and there is certainly no 
indication of a tendency for our calculational procedure 
to overestimate barrier heights.

Nevertheless, we have not yet discussed triaxiality, which could be
expected to lower fission barriers to some extent. Several studies of this 
question have been made, the most extensive probably being that of 
Ref. \cite{dutt00} (see that paper for other references), in which the 
barriers of 15 heavy and superheavy nuclei, some
close to the stability line, others highly neutron-rich, were investigated
within the framework of the ETFSI method and found, when triaxiality was
taken into account, to be lowered by 0.6 MeV on average, with a maximum shift
of 1.3 MeV. It was concluded that this effect is probably negligible, given 
the overall discrepancy between calculation and        
experiment (see Section 1). Nevertheless, two counter-examples in both of which
triaxiality is claimed to lower the barriers by the 
enormous amount of 4 MeV are to be
noted: $^{310}126$ \cite{cwiok96} and $^{258}$Fm \cite{ben98}. The latter case,
which was regrettably overlooked in Ref. \cite{dutt00}, is particularly
disturbing, since it occurs in a region of the nuclear chart where the barriers
of neighbouring nuclei are much less sensitive to triaxiality. The former of 
these two nuclei is too heavy for us to check with our codes, but we have
repeated the latter case using the ETFSI method, and find that triaxiality
lowers the barrier by 1.6 MeV. This is a significantly smaller effect than the
4.1 MeV claimed in Ref. \cite{ben98}, but it is still the largest
triaxiality shift that we have found, and really too large to be neglected. 
One is thus
forced to the conclusion that while {\it in most cases} triaxiality can
reasonably be neglected, as far as barriers are concerned, there will be
isolated cases where this is not possible. The situation is most
unsatisfactory, since such cases can only be identified by first performing
the full triaxial calculation, and doing this for the nearly 2000 nuclei
considered here would be prohibitively time-consuming. It seems that the
best that one can do at the present time is to proceed as we have done here,
neglecting triaxiality, and bear in mind that a few of our barriers, including
some low enough to be relevant to the r-process, will be considerably
overestimated because of this approximation. However, the barriers of our 
proton-deficient nuclei in the vicinity of $N$ = 184 will remain much 
higher than previously believed.

\noindent {\bf Acknowledgements}

We are indebted to B. Lorazo of the Computing Centre of the Universit\'e
de Montr\'eal for his help in facilitating our computations, and to
M. Farine for calculating the surface properties of the SkSC forces. 
P. M\"{o}ller is thanked for valuable communications.
J. M. P. acknowledges the
financial support of the FNRS (Belgium) that made possible an
extended visit to the Universit\'e Libre de Bruxelles, and
is grateful for the hospitality extended to him at the 
Institut d'Astronomie et d'Astrophysique by Prof. M. Arnould.
M. R. is a Research Associate of the FNRS.

\end{document}